\documentclass[10pt]{IEEEtran}

\usepackage{amsmath,amssymb}

\usepackage{subfigure}
\usepackage{graphicx}
\usepackage{float}
\usepackage{caption}
\usepackage{slashbox}%绘制表格斜线
\usepackage{longtable}
\usepackage{rotating}
\usepackage{multirow}
\usepackage{array}
\usepackage{bm}
\usepackage{algorithmic}

\usepackage[ruled,vlined,linesnumbered]{algorithm2e}
\usepackage{setspace}
\usepackage[usenames]{color}
\usepackage{cases}
\usepackage{CJK}
\usepackage{tabularx}
\usepackage{xcolor}%定义了一些颜色
\usepackage{colortbl,booktabs}%第二个包定义了几个*rule
\usepackage{amscd}
\usepackage{multirow}
\usepackage{bigdelim}
\usepackage{caption}
\usepackage{ragged2e}
\usepackage{indentfirst}
\usepackage{setspace}
\usepackage{epstopdf}

\definecolor{mygray}{gray}{.9}
\definecolor{myblue}{RGB}{135,206,250}
\definecolor{mybluegray}{RGB}{119,136,153}

\begin{document}
\title{Block Access Control in Wireless Blockchain Network: Design, Modeling and Analysis}
%\title{Wireless Blockchain Network: Performance Analysis of CSMA/CA with Forking Solution Scheme}

\author{Yixin Li, Bin Cao$^*$, Liang Liang, Deming Mao, and Lei Zhang\vspace{-0.45cm}
\thanks{Y. Li and L. Liang are with the School of Microelectronics and Communication Engineering, Chongqing University, Chongqing 400044, China. B. Cao (the corresponding author: \texttt{caobin@bupt.edu.cn}) is with the State Key Laboratory of Networking and Switching Technology, Beijing University of Posts and Telecommunications, Beijing 100876, China. D. Mao is with the China Electronic Technology Cyber Security Co., Ltd, Chengdu 610041, China. L. Zhang is with the James Watt School of Engineering, University of Glasgow, Glasgow, G12 8QQ, U.K. }
\thanks{This work was supported by National Natural Science Foundation of China under Grant 62071075.}}

\IEEEtitleabstractindextext{%
\begin{abstract}
\justifying
%With the evolution of wireless networks, the computing and storage resources have been deployed from remote cloud to edge. Meanwhile, to improve communication efficiency and coverage, the network structure is also changing from macrocell to small-cell. These trends result in the distribution or semi-distribution in wireless networks.
Wireless blockchain network is proposed to enable a decentralized and safe wireless networks for various blockchain applications. To achieve blockchain consensus in wireless network, one of the important steps is to broadcast new block using wireless channel. Under wireless network protocols, the block transmitting will be affected significantly. In this work, we focus on the consensus process in blockchain-based wireless local area network (B-WLAN) by investigating the impact of the media access control (MAC) protocol, CSMA/CA. With the randomness of the backoff counter in CSMA/CA, it is possible for latter blocks to catch up or outpace the earlier one, which complicates blockchain forking problem. In view of this, we propose mining strategies to pause mining for reducing the forking probability, and a discard strategy to remove the forking blocks that already exist in CSMA/CA backoff procedure. Based on the proposed strategies, we design Block Access Control (BAC) approaches to effectively schedule block mining and transmitting for improving the performance of B-WLAN. Then, Markov chain models are presented to conduct performance analysis in B-WLAN. The results show that BAC approaches can help the network to achieve a high transaction throughput while improving block utilization and saving computational power. Meanwhile, the trade-off between transaction throughput and block utilization is demonstrated, which can act as a guidance for practical deployment of blockchain.
\end{abstract}

\begin{IEEEkeywords}
Blockchain, wireless network, CSMA/CA, forking, Markov chain, performance analysis.
\end{IEEEkeywords}}

\maketitle

\IEEEdisplaynontitleabstractindextext

% \fi
\IEEEpeerreviewmaketitle



%\IEEEraisesectionheading{\section{Introduction}\label{sec:introduction}}

\section{Introduction}

Wireless blockchain network is proposed to enable a robustness and distributed wireless network for different blockchain applications, such as blockchain-based mobile edge computing \cite{1-bMEC,2-bMEC}, blockchain for vehicles management \cite{1-bvehicles,2-bvehicles}, and blockchain for smart factory \cite{2-factory}. Using blockchain to build distributed wireless networks has the following advantages: 1) Alleviate the pressure of high-load nodes in the network and reduce the impact of single point of failure. 2) Improve the security and scalability of network, and reduce maintenance cost, especially for large scale scenarios such as Internet of Things (IoT) \cite{6-IoTBC}. 3) Achieve adaptive matching and behavioral decision-making of users/terminals by involving smart contract \cite{2-smart,3-smart}.

The decentralization and security provided by blockchain for network can be largely attributed to the use of consensus algorithm \cite{6-IoTBC}, which motivates the nodes in the network to maintain a single version of the digital ledger without the involvement of a third party. Several consensus algorithms have been proposed, e.g., proof-of-work (PoW) \cite{1-bitcoin}, proof-of-stake (PoS) \cite{3-PoS}, practical Byzantine fault tolerant (PBFT) \cite{5-PBFT}, and Raft \cite{6-raftWBN}. Among them, PoW is the first widely used one in blockchain, and has a better security and node scalability than PBFT and Raft, since the fault tolerance and communication efficiency in PoW is higher than that in PBFT and Raft \cite{6-BCSecurity,6-BCscalability}. In view of these advantages, this work studies the PoW consensus process in wireless network and the analysis can be extended to PoS easily. Using PoW to achieve consensus in wireless network, one of the important steps is to broadcast new block using wireless channel. Under wireless network protocols, the block transmitting will be affected significantly.

Considering the characteristics of wireless network, this work focuses on the consensus process in blockchain-based wireless local area network (B-WLAN) by investigating the impact of carrier sense multiple access with collision avoidance (CSMA/CA), which is a random access mechanism on media access control (MAC) layer. In a typical blockchain design, with ideal communication conditions, the first full node (FN) \cite{3-Mastering} which generates a valid new block is the winner of bonus. However, as shown in Fig. 1(b), due to the randomness of the backoff counter in CSMA/CA, the first block generated by a FN may not be transmitted immediately, thus the other FNs will keep on mining to generate new blocks. In this case, more than one blocks might be generated during a backoff counter and the latter block has the probability to outpace the first one, thus become the final winner. Meanwhile, the first block will become a fork in the blockchain ledger. The forking problem results in the inconsistency of blockchain ledgers among the FNs, then lead to the waste of computational power and security issues, such as ``double-spending" \cite{8-double-spend}. Due to forking problem, the block generation rate in PoW are slowed down by blockchain system codes and thus the transaction throughput are limited to dozens usually, e.g., 7 transaction per second (tps) in Bitcoin \cite{1-bitcoin} and 15 tps in Ethereum \cite{6-ethereum}.

%Intuitively, since the average CSMA/CA backoff duration is very short, the probability of forking during the backoff couner will be low.

Furthermore, with the evolution of wireless networks, the nodes of network become more dense, and to meet the needs of a surge of service requests, the block generation rate should be accelerated to improve the transaction throughput. In this case, the forking problem will become more serious, since the frequent collisions occurring on the channel prolong the backoff counter and the new block arrives very fast, which increase the forking probability considerably. To address this forking problem and improve B-WLAN performance, we propose mining strategies to reduce the forking probability, and a discard strategy to remove the forking blocks that already exist in CSMA/CA backoff procedure. Based on the proposed strategies, we design four Block Access Control (BAC) approaches to schedule block mining and transmitting effectively. Then, using Markov chain models, we carry out mathematical analysis to show how the different BAC approaches can improve the performance of a B-WLAN. The main contributions of this paper can be summarized as follows.

\begin{itemize}
\item We propose mining strategies to pause mining during the backoff and transmission of a new block, which aim to reduce the meaningless computational power consumption on forking blocks and improve block utilization.
    %We propose two mining strategies to pause mining based on channel condition and the working state of a FN respectively, which can reduce the generation of forking blocks and improve block utilization. The principle of mining strategies is to reduce the meaningless computational power consumption on forking blocks, which is important to PoW-based blockchain.
\item We propose a discard strategy to stop FNs from broadcasting forking blocks, which act as a key enabler to accelerate block generation rate and improve transaction throughput. This strategy can work in parallel with mining strategies to improve the overall performance of B-WLAN.
\item Based on the proposed strategies, we design four BAC approaches and use Markov chain models to conduct performance analysis in B-WLAN. By analysing the stationary probability of Markov chain, we derive the closed-form expressions of key performance metrics, in terms of transaction throughput, block discard rate, block utilization and mining pause probability.
\item Our experimental results validate the effectiveness of mining strategies and discard strategy on improving the transaction throughput and block utilization of B-WLAN. We also make various interesting observations about the impact of blockchain system parameters on the performance trade-off.

%Through extensive experiments, we validate our analysis and obtain some insightful results: (i) the WBN can reach a high transaction throughput when using BAC approach to address forking problem. (ii) the mining strategy in BAC approach can save computational power and improve the block utilization of WBN, especially when the block generation rate is high. (iii) the trade-off between transaction throughput and block utilization, i.e., a high transaction throughput can be obtained at the expense of block utilization.
\end{itemize}

The rest of this paper is organized as follows. Section II describes the consensus process in B-WLAN and forking problem. Section III proposes strategies to address forking problem and improve B-WLAN performance. Based on the strategies, four BAC approaches are designed. Section IV introduces Markov chain models to capture the working process of BAC approaches. Based on the stationary probabilities of the models, the closed-form expressions of key performance metrics are analysed in Section V. Section VI conducts some experiments to compare the performance of four BAC approaches. Then, related works are discussed in Section VII. Finally, we concludes the whole paper in Section VIII.

\section{Preliminaries}

In this section, we start by introducing the main elements in B-WLAN. Then, we discuss the consensus process and forking problem in the network.

%Considering the importance of wireless communication in consensus process, we use CSMA/CA as the MAC protocol to illustrate how to broadcast a block in a B-WLAN. Meanwhile, a strategy that operates in parallel with CSMA/CA is designed to address forking in this network.

\subsection{Blockchain-Based Wireless Local Area Network}

\begin{figure}[t]
\setlength{\abovecaptionskip}{0.cm}
\setlength{\belowcaptionskip}{-0.4cm}
\captionsetup{font={footnotesize}}
\centering
\subfigure[An illustration of blockchain-based wireless local area network.]{
\includegraphics[width=6.5cm]{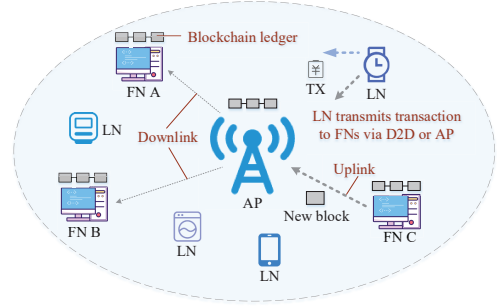}
}
\subfigure[Channel contention for block transmission.]{
\includegraphics[width=6.5cm]{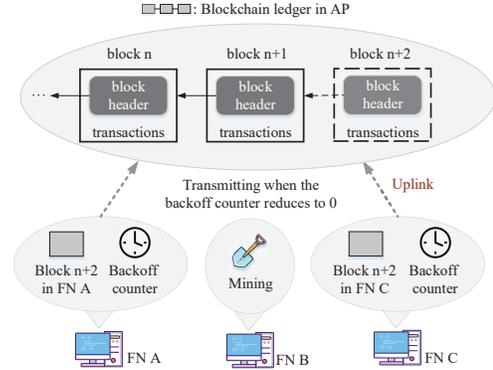}
}
\caption{{An illustration of B-WLAN and channel contention process.}}
\label{figUser}
\end{figure}

As shown in Fig. 1(a), a typical B-WLAN mainly consists of three elements: lightweight nodes (LNs), full nodes (FNs) and access point (AP). LNs are storage and power-constrained nodes, and the blockchain network allows them to issue transactions without storing the full blockchain ledger \cite{3-Mastering}. FNs are nodes with enough computing power and storage space, which perform hash operations to generate new blocks for recording transactions. FNs store a full blockchain ledger \cite{3-Mastering}. For example, in autonomous vehicles application \cite{1-bvehicles}, LNs can be the vehicles that perform local tasks, e.g., machine learning. FNs can be vehicle miners that generate new block to collect and share information, e.g., local model updates. In smart factory application \cite{2-factory}, LNs can be the factory devices for data collecting. FNs can be the factory computers for data processing and consensus achieving in blockchain.

In blockchain network, the new block of a FN should be broadcast to all the other FNs for achieving consensus. Considering the uncertainty of the geographical distribution of FNs in wireless network, it is difficult to establish device-to-device (D2D) connections for all FNs. In view of this, we consider that AP is responsible for the broadcast of new blocks in B-WLAN, which is more feasible and has a higher block broadcast efficiency than the multi-hop D2D transmission. As shown in Fig. 1(a), when a new block is generated by FN C, it transmits the block to AP through the uplink, and then AP broadcasts the block to all the FNs within its coverage radius through the downlink. During the block transmission process, AP can store the latest blocks as a backup for download. To enhance the security, one can use more APs to store and broadcast blocks redundantly. However, as a starting point, this work studies the block transmission within a single AP and assumes that the attacker cannot violate communication protocols.

For transactions, the LNs with limited power can transmit transactions to nearby trusted FNs using D2D connections, which requires less energy for transmission. Note that the transactions will be included in a block which broadcast by FNs finally, so the LNs with limited power do not need to broadcast transactions to all FNs from the start. However, if LNs do not trust nearby FNs or the D2D connection is not supported in the environment, LNs can broadcast transactions via AP with a higher cost and longer delay. Since this work focuses on the block transmission, we do not further address the wireless link selection for transactions.

As we know, IEEE 802.11 series is a cost-efficient solution for WLAN that can satisfy most communication requirements in domestic, public, and business scenarios \cite{6-DCF}. The primary MAC protocol of IEEE 802.11 is CSMA/CA, which is called distributed coordination function (DCF).
In this work, we consider CSMA/CA as the access mechanism for block transmission from FN to AP, and proposes strategies to address forking during block transmission process. Nevertheless, the proposed strategies can work in parallel with the other wireless network protocols in a similar manner. At each block transmission with CSMA/CA, the backoff time is uniformly chosen in the range $[0,W_i-1]$, where $W_i$ is the contention window and $W_i = 2^{i}W_{min}$ ($0 \leq  i \leq  m$). The value of $W_i$ depends on the backoff stage $i$ (the number of retransmissions). At the first transmission attempt, contention window is equal to the minimum contention window $W_{min}$. After each unsuccessful transmission, $W_i$ is doubled, up to a maximum value $W_{max}$. We denote $m$ as the maximum number of retransmissions, where $W_m \leq  W_{max}$. A block will be discarded when $m$-th transmission is unsuccessful.

\subsection{Consensus Process and Blockchain Forking}

The main step of the consensus process in B-WLAN can be summarized as follows: 1) The LNs generate transactions and broadcast them to FNs through the wireless link. 2) All FNs collect the new transactions and perform hash operations to generate a valid new block. 3) The FN which has a valid new block competes with other FNs based on CSMA/CA for transmitting its block to AP. 4) AP receives and checks the new block, and then broadcasts it to all FNs. 5) The other FNs receive and check the new block, then insert it into their local ledgers. If any FN does not receive the new block, it can download block from AP. 6) When AP and most of the FNs have the identical copy of the new block in their local ledgers, the new block and the transactions included in it achieve preliminary consensus successfully. 7) After that, with the accumulating of blocks sequentially, the cost of attack and malicious modification will be increased exponentially \cite{1-bitcoin}.

In blockchain consensus process, due to communication delay, more than one blocks at the same height (the position in blockchain) might be created by different FNs, which result in forking problem. To describe the forking problem, we define three working states of a FN: no block, block backoff and block transmitting. Based on the definition, the forking problem will occur when a new block is generated by a FN while the other FNs are in the block backoff or block transmitting states.

\section{Block Access Control}

In this section, we propose mining strategy and discard strategy to address forking problem and improve the performance of B-WLAN. Based on the proposed strategies, we design four BAC approaches to schedule block mining and transmitting.

%to meet the performance requirement of different scenarios.

%which is a key enabler to improve the transaction throughput and block utilization of WBN

\subsection{Forking Solution}

\begin{bfseries}Mining strategy\end{bfseries}: The principle of this strategy is to pause mining (hash operations) during the backoff and transmission of a new block, which aims to reduce forking probability and improve the block utilization. %(defined as the ratio of the main chain blocks to the total generated blocks).
Specifically, there are two strategies to pause mining as follows, where strategy I pauses mining based on the block transmission detected on the channel and strategy II pauses mining based on the working state of a FN.

\emph{Strategy I:} the mining of a FN should be paused whenever a block transmission of the other FN is detected on the channel. The FN resumes the mining when the channel is detected as idle more than a distributed inter-frame space (DIFS)\footnote{If a FN generates a block during a DIFS, it can not start the backoff procedure until the end of the DIFS, which increases the forking probability.}. The reason to pause mining in this case is that when a new block of the other FN is transmitted on the channel, the current mining with the hash of an old block will generate forking blocks or waste computational power. To implement this strategy, an option is to set an additional Flag field in packet header to announce the block transmission. Accordingly, the FN can distinguish the block transmission to pause mining. Meanwhile, the time to pause mining can be easily determined based on the Duration/ID field in packet header, which contains the data transmission time to update the network allocation vector (NAV) in CSMA/CA \cite{3-performance}. In reality, there are multiple wireless access channels under a AP. In this case, AP can announce the block transmission to all the FNs using the CTS frame in CSMA/CA.

\emph{Strategy II:} the mining of a FN should be paused when the FN generates a new block. The FN resumes mining when the new block is transmitted successfully or discarded. The reason is that a new block might be overtaken by another block with the same height during the random backoff counter, especially when the block generation rate is high. So if a FN in the block backoff state keeps mining, it will generate forking blocks or waste computational power.

Although the mining strategy can reduce the forking probability, it cannot solve the forking problem in B-WLAN thoroughly, since a FN in wireless network cannot detect whether the other FNs are in the block backoff state. In view of this, we propose a discard strategy as follows.

\begin{bfseries}Discard strategy\end{bfseries}: The principle of this strategy is to discard forking blocks before they are broadcast to the network. Specifically, if a FN receives a same height block transmitted by the other FN, it should discard its own blocks that do not be broadcast yet. Actually, blocks will be discarded in the following two cases: (i) a FN in the no block state generates new blocks while a successful block transmission occurs on the channel. (ii) a FN in the block backoff state receives a block of the other FN. Using this strategy, the forking blocks will not be broadcast to the network, and thus the hash difficulty of PoW can be very low for accelerating block generation rate and improving transaction throughput.

%Although the mining strategy can reduce the forking probability, it cannot solve the forking problem in WBN thoroughly, due to a FN cannot detect whether the other FNs are in the block backoff state. To solve this problem, we propose a discard strategy to remove the forking blocks that already generated by FNs. Using this strategy, the forking blocks will not be broadcast to the network, and thus the hash difficulty of PoW can be very low for accelerating block generation rate and improving transaction throughput.

%Although the mining strategy can reduce the forking probability, it cannot solve the forking problem in WBN thoroughly, due to a FN cannot detect whether the other FNs are in the block backoff state. To solve this problem, we propose a discard strategy to remove the forking blocks that already generated by FNs. Specifically, if a FN receives a same height block transmitted by the other FN, it should discard its own blocks that do not be broadcast yet. Actually, a block will be discarded in the following two cases: (i) a FN in the no block state generates a new block while a successful block transmission occurs on the channel. (ii) a FN in the backoff state receives a block of other FN. Using this strategy, the forking blocks will not be broadcast to the network, and thus the hash difficulty of PoW can be very low for accelerating block generation rate and improving transaction throughput.

\newcommand{\tabincell}[2]{\begin{tabular}{@{}#1@{}}#2\end{tabular}}
\renewcommand\arraystretch{2}
\begin{table}[!t]
\centering
\scriptsize
\caption{Four BAC approaches}
%\begin{tabular}{%
%>{\Columncolor{Mygray}}C|%
%>{\Columncolor{White}}L|%
%>{\Columncolor{Mygray}}L|%
%>{\Columncolor{White}}L|%
%>{\Columncolor{Mygray}}L}
\begin{tabular}{|
c|%
c|%
c|%
c|%
c|
c|}

\arrayrulecolor{mybluegray}\hline
%\rowcolor{mygray}

  & Discard strategy & Mining strategy I & Mining strategy II \\
\arrayrulecolor{mybluegray}\hline
\tabincell{l}{BAC-1} & \tabincell{l}{ \checkmark }
                                    & \tabincell{l}{-}
                                    & \tabincell{l}{-} \\

\arrayrulecolor{mybluegray}\hline
\tabincell{l}{BAC-2}   & \tabincell{l}{\checkmark}
            & \tabincell{l}{ \checkmark}
            & \tabincell{l}{-}\\

\arrayrulecolor{mybluegray}\hline
\tabincell{l}{BAC-3}   & \tabincell{l}{\checkmark}
            & \tabincell{l}{-}
            & \tabincell{l}{\checkmark}\\

\arrayrulecolor{mybluegray}\hline
\tabincell{l}{BAC-4}   & \tabincell{l}{\checkmark}
            & \tabincell{l}{\checkmark}
            & \tabincell{l}{\checkmark}\\

\arrayrulecolor{mybluegray}\hline

\end{tabular}
\end{table}

\subsection{Working Approaches}

Based on the proposed strategies, we design four BAC approaches to schedule block mining and transmitting. As shown in Table I, BAC-1 only contains discard strategy; BAC-2 contains discard strategy and mining strategy I, BAC-3 contains discard strategy and mining strategy II, BAC-4 contains all the strategies.

\begin{figure}[t]
\setlength{\abovecaptionskip}{0.cm}
\setlength{\belowcaptionskip}{-0.3cm}
\captionsetup{font={footnotesize}}
\begin{center}
 \includegraphics[width=8.5cm]{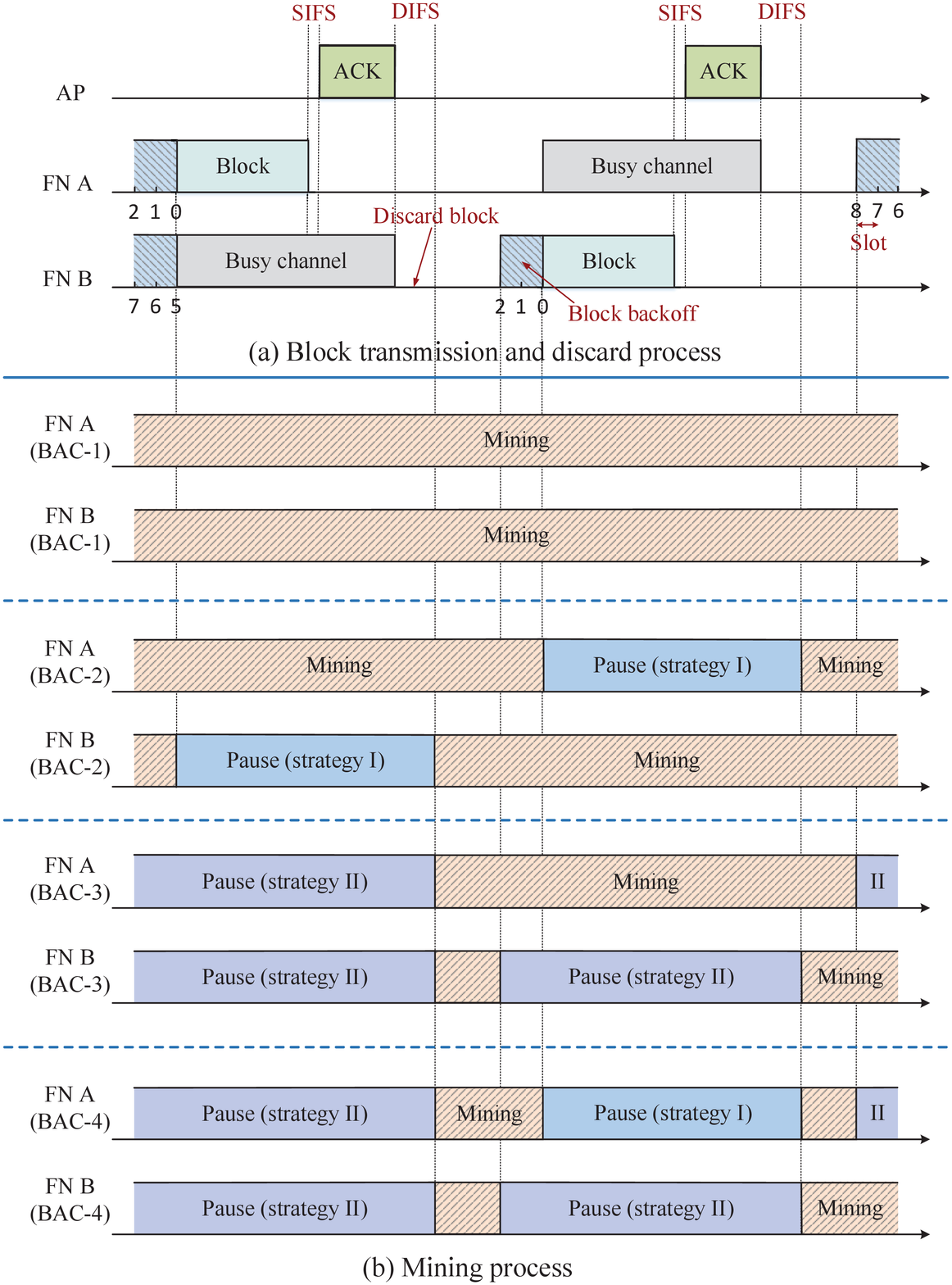}
 \end{center}
 \caption{{An example of BAC approaches to work in parallel with CSMA/CA.}}
\label{Fig2}
\end{figure}

Fig. \ref{Fig2} shows an example of how BAC approaches can work in parallel with CSMA/CA. In this example, we suppose the block is generated simultaneously in four BAC approaches to show when a FN should discard block and pause mining. Accordingly, the block transmission and discard process in Fig. \ref{Fig2}(a) are the same in four BAC approaches, since all the approaches contain discard strategy and use CSMA/CA to transmit blocks. We can see that a backoff block of FN B is discarded after the block transmission of FN A is successful.
%Here, we consider that the average downlink broadcast delay of AP is not longer than DIFS.
On the other hand, the mining process in Fig. \ref{Fig2}(b) is different among four BAC approaches, which is controlled by mining strategy I and strategy II.

\begin{bfseries}BAC-1\end{bfseries}: only contains discard strategy, thus a FN will keep mining all the time, which is shown in Fig. \ref{Fig2}(b). In no block state, if the FN generates a new block while a successful block transmission occurs on the channel, the FN discards its own block based on the discard strategy. If the new block is generated during the other information transmission, collision or channel idle time, the FN schedules its block transmission based on CSMA/CA. Once the channel remains idle more than a DIFS, the FN starts the backoff procedure by selecting a random initial value as the backoff counter. In block backoff state, the FN decreases the backoff counter while listening to the channel. Whenever a block transmission is detected on the channel, the backoff counter is paused and the FN begins to receive a new block. If the block transmission is successful, the FN discards its own block. Otherwise, the FN discards the collision message and continues to listen to the channel. If the channel remains idle more than a DIFS, the backoff counter is resumed. When the backoff counter reaches zero, the FN enters block transmitting state. If no other blocks and information transmit at this time, the block transmission is successful. Otherwise, the backoff stage increases by one and the FN stays in the backoff state.

Since a FN using the BAC-1 keeps mining all the time, it can generate new blocks during the block backoff and block transmitting states, which results in the queueing of new blocks. In this case, once a FN receives a new block of the other FN, all the blocks in the queue should be discarded. The reason is that the block in queue references the hash of the backoff block, and the discard of a earlier block will invalidate the following blocks in blockchain. Another impact of the mining during backoff and transmitting states is that, if there exist blocks in queue, a FN can directly schedule a new transmission after the earlier transmission is finished.

\begin{bfseries}BAC-2\end{bfseries}: contains discard strategy and mining strategy I, thus the mining of a FN will be paused whenever a block transmission of the other FN is detected on the channel until the channel is idle more than a DIFS, which is shown in Fig. \ref{Fig2}(b). Due to the mining pause, the expected mining time of FN in BAC-2 is less than that in BAC-1, which results in two differences between BAC-2 and BAC-1. The first is that, for a randomly chosen time slot, the probability to leave no block state in BAC-2 is lower than that in BAC-1. The second is that the expected time to generate queueing blocks in BAC-2 is less than that in BAC-1.

%Since the mining is paused, it is impossible for FN to generate blocks during the block transmission. As a result, when a FN is in no block state, the probability to generate new block using BAC-2 is lower than that using BAC-1. Meanwhile, after a FN transmits or discards its block, the expected number of queueing blocks in BAC-2 will be less than that BAC-1, due to
%
%Except the probability to generate new block in no block state and the expected number of queueing blocks, the other behaviors of BAC-2 is similar to BAC-1.

\begin{bfseries}BAC-3\end{bfseries}: contains discard strategy and mining strategy II, thus the mining of a FN will be paused when the FN has a new block until the new block is transmitted successfully or discarded, which is shown in Fig. \ref{Fig2}(b). Since the mining is paused during the block backoff and block transmitting states, block queueing does not exist in BAC-3. So after a new block is transmitted successfully or discarded, a FN will return to no block state and restart mining. Except that BAC-3 has no block queueing, the other behaviors of BAC-3 is similar to BAC-1.

\begin{bfseries}BAC-4\end{bfseries}: contains all the strategies. There are two differences between BAC-4 and BAC-1. The first is that, for a randomly chosen time slot, the probability to leave no block state in BAC-4 is lower than that in BAC-1. The second is that the block queueing does not exist in BAC-4, thus a FN will return to no block state and restart mining after the FN transmits or discards its block.

\section{Mathematical Modelling}

In this section, we formulate the working process of four BAC approaches as Markov chain models to study the stationary probabilities, which act as the basis for performance analysis.

\subsection{Markov Chain Model for BAC-1 and BAC-2}

In this work, we analyse the maximum transaction throughput in B-WLAN by assuming that an independent channel is assigned for uplink block transmission. Based on this assumption, the other information is not interfere with block transmission, and we only consider the competition of blocks in the following analysis. Another assumption is that the channel condition is ideal \cite{3-performance}, i.e., the only reason of a transmission failure is that a collision occurs on the channel.

%We assume that the average broadcast delay of downlink is not longer than DIFS. Based on the assumption, the backoff blocks will be discarded by FNs during the DIFS, and thus all FNs can restart mining based on new block once the DIFS ends.

We consider a B-WLAN has $N$ FNs. Each FN has three working states: no block, block backoff and block transmitting, which can be depicted as Fig. 3. In this model, the no block state is described by $\{-1,0\}$; the block backoff states are described by $\{i,k\}$, where $i \in  [0,m]$ representing the backoff stage and $k \in [1,W_{i}-1]$ representing the backoff counter in time slots; the block transmitting states are described by $\{i,0\}$ ($i \in  [0,m]$), in which the block will be transmitted to channel. When a given FN is in the no block state or backoff state, the channel with probability $p_{s}$ contains a successful block transmission; the channel with probability $p_{c}$ contains a collision; the channel with probability $1 -p_{s} -p_{c}$ stays idle. Different with this, when a given FN is in the transmitting state, its block will be transmitted to channel, and thus a successful block transmission occurs on the channel with probability $1 -p_{s} -p_{c}$; a collision happens with probability $p_{s} +p_{c}$. Let $T_{s}$ be the average channel busy time when a successful transmission occurs on the channel, $T_{c}$ be the average channel busy time when a collision happens, and $\sigma$ be the size of time slot. Similar with \cite{3-performance}, the Markov chain model adopts a discrete time scale, and one step in this model can be $T_{s}$, $T_{c}$ or $\sigma$, which is determined by the channel condition, i.e., the channel may contain a successful transmission, a collision or stay idle.

\begin{figure}[t]
\setlength{\abovecaptionskip}{0.cm}
\setlength{\belowcaptionskip}{-0.3cm}
\captionsetup{font={footnotesize}}
\begin{center}
 \includegraphics[width=8.5cm]{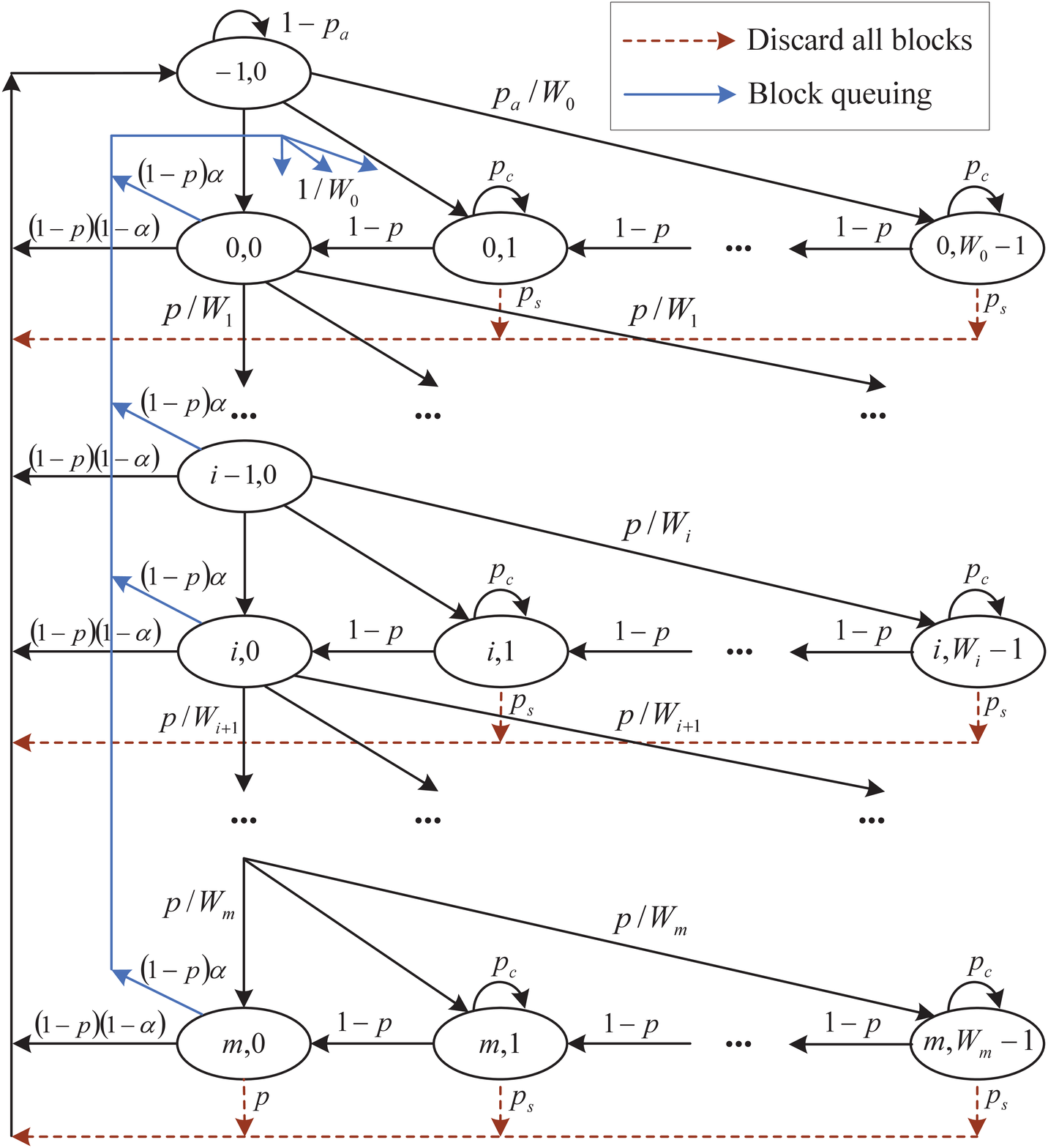}
 \end{center}
 \caption{Markov chain model for BAC-1; When ${p}_{a}$ and $\alpha$ change to $\tilde{p}_{a}$ and $\tilde{\alpha}$ respectively, this model is available for BAC-2.}
\label{Basic}
\end{figure}

\begin{bfseries}BAC-1\end{bfseries}: Using the BAC-1, a FN will perform hash operations with hashrate $r$ (the number of hash operations per second) all the time. Let the hash difficulty of PoW in this network be $D$ representing the expected number of hash operations to find a valid block. Based on \cite{3-PoS}, we have the time $T$ for a FN to find a valid block is exponentially distributed with block generation rate $\lambda = r/D$. Now we study the behavior of a FN in the no block state and let ${p}_{a}$ be the probability that a FN generates a new block and leaves no block state during a step of Markov chain model. In no block state, if the channel contains a successful block transmission, the FN will stay no block state during this step no matter whether the FN generates a new block or not, since all the blocks that do not be broadcast will be discarded by the discard strategy; if the channel contains a collision, the mining time of the FN in this step will be $T_{c}$ and thus the probability to leave no block state will be $P\{T \leq T_{c}\}$; if the channel stays idle, the mining time will be $\sigma$ and thus the probability to leave no block state will be $P\{T \leq \sigma\}$. In summary, we have
\begin{equation}\label{pa1}\small
\begin{split}
{p}_{a}=&{p}_{s}\cdot 0+{p}_{c}\cdot P\{T\leq T_{c}\}+(1-{p}_{s}-{p}_{c})\cdot P\{T\leq\sigma\}\\
=&{p}_{c}[1-\exp (-\lambda T_{c})]+(1-{p}_{s}-{p}_{c})[1-\exp (-\lambda\sigma)].
\end{split}
\end{equation}
Acoordingly, the one-step transition probabilities in no block state can be given by
\begin{equation}\label{onestep22}\small
\begin{split}
\begin{cases}
P\left\{ {-1,0\mid -1,0} \right\} = 1-{p}_{a},\\
P\left\{ {0,k\mid -1,0} \right\} = {p}_{a}/W_0,~~~~~k\in[0,W_0 - 1].
\end{cases}
\end{split}
\end{equation}

The FN leaves the no block state with probability $p_{a}$ and uniformly chooses a backoff time from $[0,W_0 - 1]$. Then, it starts decreasing the backoff counter while listening to the channel. In block backoff states, the FN discards its blocks with probability ${p}_{s}$; the FN pauses the backoff counter with probability ${p}_{c}$; the FN decreases the backoff counter with probability $1 - {p}$ ($p = p_{s} + p_{c}$). Thus, the one-step transition probabilities in block backoff states can be given by
\begin{equation}\label{onestep21}\small
\begin{split}
\begin{cases}
P\left\{ {-1,0 \mid i,k} \right\} = {p}_{s},~~~~~ \, ~~i\in[0,m],\,k\in[1,W_i-1],\\
P\left\{ {i,k \mid i,k} \right\} = {p}_{c},~~~~~~~~~~i\in[0,m],\,k\in[1,W_i-1],\\
P\left\{ {i,k - 1 \mid i,k} \right\} = 1 - {p},~i\in[0,m],\,k\in[1,W_i-1].
\end{cases}
\end{split}
\end{equation}

When the backoff counter becomes $0$, the FN will enter the block transmitting state $\{i,0\}$ and its block will be transmitted into the channel. If no other FNs transmit block in this slot, the transmission is successful. Otherwise, the backoff stage increases and the FN starts a new backoff counter. Especially, in state $\{m,0\}$, the FN will discard its blocks when a collision occurs. Let $p$ be the probability that a collision is seen by a block transmitted on the channel, where $p = p_{s} +p_{c}$.

%In this model, the conditional collision probability $p$ is assumed to be constant and independent with the number of collisions already suffered \cite{3-performance}.

Now we study the one-step transition probabilities in block transmitting states. Since a FN using the BAC-1 keeps mining all the time, it may generate queueing blocks during the backoff and transmitting states. In this case, when a block discard occurs in block backoff states or state $\{m,0\}$, all the queueing blocks will also be discarded and thus the FN returns to no block state, shown as the dotted lines in Fig. \ref{Basic}. On the other hand, if there exist queueing blocks after a successful transmission in a block transmitting state, a FN will directly move from the block transmitting state to the block backoff state, shown as the blue lines in Fig. \ref{Basic}. Let $\alpha$ be the stationary probability that the queue is not empty after a successful block transmission. Based on \cite{14-Probability}, $\alpha = \lambda T_{q}$, where $\lambda$ is the block generation rate of a FN, and $T_{q}$ is defined as the expected time of a block spent on backoff and transmitting states counting from a block generation to its successful transmission. Using the expectation formula, $T_{q}$ can be given by
\begin{equation}\label{Tq}
T_{q} =  \sum\limits_{i = 0}^m {p_e}(i)\left[(i{T_c}  +  {T_s}) + {\sum\limits_{n = 0}^i {\frac{{{W_n}  -  1}}{2}} }\left(\sigma + \frac{{p}_{c}}{1 - {p}}T_{c}\right)\right],
\end{equation}
where $p_{e}(i)$ ($i \in [0,m]$) is the probability that a block exits the backoff scheme through a successful transmission in state $\{i,0\}$, and it can be expressed as
\begin{equation}\label{pe}
{p_e}(i) = \prod\limits_{n = 0}^i {\frac{{1 - {{\left(\frac{1 - {p}}{1 -{{p}_c}}\right)}^{{W_n}}}}}{{{W_n}}}}\left[\frac{{p}}{{1 - \frac{1 - {p}}{1 -{{p}_c}}}}\right ]^{i+ 1}\frac{1 - {p}}{{p}}.
\end{equation}
The complete proof of $T_{q}$ in (\ref{Tq}) is given in the Appendix. Based on (\ref{Tq}), $\alpha$ can be given by
\begin{equation}\label{alpha}
\alpha = \lambda \sum\limits_{i = 0}^m {p_e}(i)\left[(i{T_c}  +  {T_s}) + {\sum\limits_{n = 0}^i {\frac{{{W_n}  -  1}}{2}} }\left(\sigma + \frac{{p}_{c}}{1 - {p}}T_{c}\right)\right].
\end{equation}

Once $\alpha$ is determined, the one-step transition probabilities in transmitting states are
\begin{equation}\label{onestep23}\small
\begin{split}
\begin{cases}
P\left\{ {0,k \mid i,0} \right\}  =  (1 - {p})\alpha/W_{0},~~~i \in [0,m],~ ~~~~k \in [0,W_0 - 1],\\
P\left\{ {i + 1,k \mid i,0} \right\}  =  {p}/W_{i+1},~~~~~i \in [0,m-1],k \in [0,W_0 - 1],\\
P\left\{ {-1,0 \mid i,0} \right\}  = (1 - {p})(1 - \alpha),i \in [0,m-1],\\
P\left\{ {-1,0 \mid m,0} \right\}  =  (1 - {p})(1 - \alpha) + {p}.
\end{cases}
\end{split}
\end{equation}

\begin{bfseries}BAC-2\end{bfseries}: The mining pause during the block transmission results in two differences between BAC-2 and BAC-1. The first is the probability to leave no block state during a step of Markov chain model. The second is the expected time to generate queueing blocks.

Let $\tilde{p}_{a}$ be the probability that a FN using BAC-2 generates a new block and leaves no block state during a step of Markov chain model. Compared with (\ref{pa1}) in BAC-1, the FN using BAC-2 will pause mining whenever a block transmission of the other FN is detected on the channel, so $\tilde{p}_{a}$ is given by
\begin{equation}\label{pa2}
\begin{split}
\tilde{p}_{a}=&{p}_{s}\cdot 0+{p}_{c}\cdot 0+(1-{p}_{s}-{p}_{c})\cdot P\{T\leq \sigma\}\\
=&(1-{p}_{s}-{p}_{c})[1-\exp (-\lambda\sigma)].
\end{split}
\end{equation}

Let $\tilde{\alpha}$ be the stationary probability that the queue is not empty after a FN using BAC-2 transmits its block successfully. Let $\tilde{T}_{q}$ be the expected time to generate queueing blocks in BAC-2. Based on the definition, we have $\tilde{\alpha}=\lambda\tilde{T}_{q}$. Using BAC-2, a FN performs hash operations when the channel is idle or when the FN transmits its own block. Accordingly, by means of (\ref{Tq}), we have \begin{equation}\label{tTq}
\tilde{T}_{q} =  \sum\limits_{i = 0}^m {p_e}(i)\left[(i{T_c}  +  {T_s}) + {\sum\limits_{n = 0}^i {\frac{{{W_n}  -  1}}{2}} }\sigma\right].
\end{equation}
Then, $\tilde{\alpha}=\lambda\tilde{T}_{q}$ can be derived. When ${p}_{a}$ and $\alpha$ change to $\tilde{p}_{a}$ and $\tilde{\alpha}$ respectively, the one-step probabilities in (\ref{onestep22}), (\ref{onestep21}) and (\ref{onestep23}) are available for BAC-2.

\subsection{Stationary Probabilities for BAC-1 and BAC-2}

Let $\pi_{-1,0}$ denotes the stationary probability of no block state, $\pi_{i,k}$ ($i \in [0,m],\,k \in [0,W_i - 1]$) denote the stationary probabilities of backoff states and transmitting states. Based on the chain regularities of the backoff and transmitting states in Fig. 3, we can obtain
\begin{equation}\label{regular1}
\begin{split}
\begin{cases}
\pi _{i,0} = \frac{p}{W_i}{\pi _{i - 1,0}}  +\! (1 \!-\! p)\pi _{i,1},\\
\pi _{i,1} = \frac{p}{W_i}{\pi _{i - 1,0}} \!+\! (1 \!-\! p)\pi _{i,2}\!+\!p_{c}\!\cdot\!\pi _{i,1},\\
\pi _{i,2} = \frac{p}{W_i}{\pi _{i - 1,0}} \!+\! (1 \!-\! p)\pi _{i,3}\!+\!p_{c}\!\cdot\!\pi _{i,2},\\
\cdots,\\
\pi _{i,W_{i\!-\!1}} \!=\! \frac{p}{W_i}{\pi _{i - 1,0}}\!+\!p_{c}\!\cdot\!\pi _{i,W_{i\!-\!1}},
\end{cases}
\end{split}
\end{equation}
where $i\!\in\![1,m]$. Using the second equation of (\ref{regular1}), $\pi _{i,0}$ can be rewritten as $\pi _{i,0} \!=\! (1\!+\!\frac{1-p}{1-p_{c}})\frac{p}{W_i}{\pi _{i - 1,0}} \!+\! \frac{({1-p})^{2}}{1-p_{c}}\pi _{i,2}$. In the same way, the other equations in (\ref{regular1}) can be used to simplify $\pi _{i,0}$ as follows:
\begin{equation}\label{i01}
\begin{split}
\pi _{i,0} \!&=\! \left[1\!+\!\frac{1\!-\!p}{1\!-\!p_{c}}\!+\!{(\frac{{1 \!-\! p}}{{1 \!-\! {p_c}}})^2}\!+\!\cdots\!+\!{(\frac{{1 \!-\! p}}{{1 \!-\! {p_c}}})^{W_{i\!-\!1}}}\right]\frac{p}{W_i}{\pi _{i - 1,0}} \\
\!&=\! \frac{{1 \!-\! {{[(1 \!-\! p)/(1 \!-\! {p_c})]}^{{W_i}}}}}{{1 \!-\! (1 \!-\! p)/(1 \!-\! {p_c})}}\frac{p}{W_i}{\pi _{i - 1,0}} ,
\end{split}
\end{equation}
where $i\!\in\![1,m]$. Based on (\ref{i01}), we obtain
\begin{equation}\label{001}
\begin{split}
{\pi} _{i,0} \!=\!{\prod\limits_{n = 0}^i {\frac{{1 \!-\! {{[(1 \!-\! {p})/(1 \!-\! {{p}_c})]}^{{W_n}}}}}{{{W_n}}}} {{\left[\frac{{p}}{{1 \!-\! (1\! -\! {p})/(1 \!-\! {{p}_c})}}\right]}^{i }}}{{\pi} _{0,0}},
\end{split}
\end{equation}
where $i\!\in\![1,m]$.

Then, we study the case when $i\!=\!0$. From Fig. \ref{Basic}, we can obtain that the regularity to enter stage $0$ is
\begin{equation}\label{regularity0}\small
{\pi} _{0,k}\!=\! \frac{1}{W_{0}}\left[{{p}_{a}}{{\pi} _{ - 1,0}} \!+\! (1 \!-\! {p})\alpha \sum\limits_{i = 0}^m {{{\pi} _{i,0}}}\right],~~k\!\in\![0,W_0\!-\!1].
\end{equation}
Meanwhile, the regularity to enter stage $i$ ($i\!\in\![1,m]$) is
\begin{equation}\label{regularityi}
{\pi} _{i,k}\!=\! \frac{1}{W_{i}}{p}{\pi} _{i-1,0},~~~~i\!\in\![1,m],k\!\in\![0,W_i\!-\!1].
\end{equation}
Based on the regularities in (\ref{regularity0}) and (\ref{regularityi}), we can use ${{p}_{a}}{{\pi }_{ - 1,0}} \!+\! (1 \!-\! {p})\alpha \sum\limits_{i = 0}^m {{{\pi} _{i,0}}}$ to replace ${p}{{\pi} _{i-1,0}}$ ($i\!\in\![1,m]$) in (\ref{i01}) and this yields
\begin{equation}\label{0011}
\begin{split}
{\pi} _{0,0} \!=\! \frac{{1 \!-\! {{[(1 \!-\! {p})/(1 \!-\! {{p}_c})]}^{{W_0}}}}}{{1 \!-\! (1 \!-\! {p})/(1 \!-\! {{p}_c})}}\frac{{p}_{a}}{W_0}\left[{\pi _{ - 1,0}} \!+\! \frac{(1 \!-\! {p})\alpha}{{p}_{a}} \sum\limits_{i = 0}^m {{{\pi} _{i,0}}}\right].
\end{split}
\end{equation}
To simplify (\ref{0011}), we should find the expression of ${\pi} _{ - 1,0}$. Based on the chain regularity of no block state in Fig. 3, we have
\begin{equation}\label{regularityno}\small
\begin{split}
{{\pi} _{ -1,0}} \!= &(1 \!-\! {{p}_{a}}){{\pi} _{ -1,0}} \!+\! (1\! -\! {p})(1 \!-\! \alpha )\sum\limits_{i = 0}^m {{{\pi} _{i,0}}}  \!+\! {p}{{\pi} _{m,0}} \!\\
&+\! {{p}_s}\left(1 \!-\! \sum\limits_{i = 0}^m {{{\pi} _{i,0}}} \! -\! {{\pi} _{ -1,0}}\right).
\end{split}
\end{equation}
After simplifying (\ref{regularityno}), we obtain
\begin{equation}\label{regularitynos}\small
\begin{split}
{{\pi} _{ -1,0}} \!= &\frac{(1\! -\! {p})(1 \!-\! \alpha )}{{p}_{a}\!+\!{p}_{s}}\sum\limits_{i = 0}^m {{{\pi} _{i,0}}}\!+\! \frac{{p}}{{p}_{a}\!+\!{p}_{s}}{{\pi} _{m,0}} \!\\
&+\! \frac{{{p}_s}}{{p}_{a}\!+\!{p}_{s}}\left(1 \!-\! \sum\limits_{i = 0}^m {{{\pi} _{i,0}}}\right).
\end{split}
\end{equation}
Substituting (\ref{001}) and (\ref{regularitynos}) into (\ref{0011}) yields
%\begin{figure*}\small
%\begin{equation}\label{002}\small
%\begin{split}
%\pi _{0,0} \!=\! \frac{{1 \!-\! {{[(1 \!-\! p)/(1 \!-\! {p_c})]}^{{W_0}}}}}{{1 \!-\! (1 \!-\! p)/(1 \!-\! {p_c})}}\frac{p_{a*}}{W_0}\left[\frac{(1\! -\! p)(1 \!-\! \alpha )}{p_{a*}\!+\!p_{s}}\sum\limits_{i = 0}^m {{\pi _{i,0}}}\!+\! \frac{p}{p_{a*}\!+\!p_{s}}{\pi _{m,0}} \!+\! \frac{{p_s}}{p_{a*}\!+\!p_{s}}(1 \!-\! \sum\limits_{i = 0}^m {{\pi _{i,0}}})\!+\! \frac{(1 \!-\! p)\alpha}{p_{a*}} \sum\limits_{i = 0}^m {{\pi _{i,0}}}\right].
%\end{split}
%\end{equation}
%\end{figure*}
%Then substituting (\ref{001}) into (\ref{002}) gives
\begin{equation}\label{003}\small
\begin{split}
{\pi} _{0,0} \!=& \left[\frac{(1\! -\! {p})(1 \!-\! \alpha )}{{p}_{a}\!+\!{p}_{s}}\!-\!\frac{{{p}_s}}{{p}_{a}\!+\!{p}_{s}}\!+\!\frac{(1 \!-\! {p})\alpha}{{p}_{a}}\right]\sum\limits_{i = 0}^m f(i){{{\pi} _{0,0}}}\!\\
&+\! \frac{{p}}{{p}_{a}\!+\!{p}_{s}}f(m){{\pi} _{0,0}}\!+\!\frac{{{p}_s}}{{p}_{a}\!+\!{p}_{s}}\frac{{1 \!-\! {{[(1 \!-\! {p})/(1 \!-\! {{p}_c})]}^{{W_0}}}}}{{1 \!-\! (1 \!-\! {p})/(1 \!-\! {{p}_c})}}\frac{{p}_{a}}{W_0}.
\end{split}
\end{equation}
where {\small{$f(x)\!=\!\!{\prod\limits_{n = 0}^x \!{\frac{{1 - {{[(1 - {p})/(1 - {{p}_c})]}^{{W_n}}}}}{{{W_n}}}} {{\left[\frac{{p}}{{1 - (1 - {p})/(1 -{{p}_c})}}\right]}^{x \!+\! 1}}} \frac{{{{p}_{a}}}}{{p}}$}}. After simplifying (\ref{003}), we obtain the expression of $\pi _{0,0}$ as follows:
\begin{equation}\label{00f}\small
\begin{split}
{\pi} _{0,0} \!=\!\frac{\frac{{{p}_{s}}}{{p}_{a}\!+{p}_{s}}\frac{{1 - {{[(1 - {p})/(1 - {{p}_{c}})]}^{{W_0}}}}}{{1 - (1 - {p})/(1 -{{p}_{c}})}}\frac{{p}_{a}}{W_0}}{1\!+\!\left[\frac{{{p}_s}}{{p}_{a}\!+{p}_{s}}\!-\!\frac{(1 - {p})(1 - \alpha )}{{p}_{a}\!+{p}_{s}}\!-\!\frac{(1 -{p})\alpha}{{p}_{a}}\right]\sum\limits_{i = 0}^m f(i)\!-\! \frac{{p}}{{p}_{a}\!+{p}_{s}}f(m)}.
\end{split}
\end{equation}

Let $\tau_{1}$ be the probability that a FN using BAC-1 transmits block in a randomly chosen time slot. It can be expressed as {\small$\tau_{1} \! =\! \sum\limits_{i = 0}^m {{\pi _{i,0}}}$}. $\tau_{b}$ rewrites as
\begin{equation}\label{taotao}\small
\begin{split}
{\tau_{1}}  \!=\! \left\{1 \!+\! \sum\limits_{i = 1}^m \!{\prod\limits_{n = 1}^i \!{\frac{{1 \!-\! {{[\frac{(1 - {p})}{(1 - {p}_{c})}]}^{{W_n}}}}}{{{W_n}}}} } {{\left[\frac{{p}}{{1 \!- \!\frac{(1 - {p})}{(1 - {p}_{c})}}}\right]}^i}\right\}{{\pi} _{0,0}},
\end{split}
\end{equation}
where ${\pi} _{0,0}$ can be substituted by (\ref{00f}), $\alpha$ can be substituted by (\ref{alpha}), ${p}_{a}$ can be substituted by (\ref{pa1}). The probability $p$ that a collision is seen by a block transmitted on the channel is given by $p\!=\!1\!-\!{(1\!-\!\tau_{1})}^{N\!-\!1}$. The probability that the channel contains a successful block transmission of other FNs is given by $p_{s}\!=\!(N\!-\!1)\tau_{1}{(1\!-\!\tau_{1})}^{N\!-\!2}$. The probability that the channel contains a block collision of other FNs is $p_{c}=p-p_{s}$. So ${\tau_{1}}$ is the only unknown parameter in (\ref{taotao}), which can be obtained through iteration method. Note that (\ref{taotao}) can be applied to BAC-2 when ${p}_{a}$ and $\alpha$ change to $\tilde{p}_{a}$ and $\tilde{\alpha}$. To distinguish this difference between BAC-1 and BAC-2, we denote the transmitting probability of BAC-2 by $\tau_{2}$.

\subsection{Markov Chain Model for BAC-3 and BAC-4}

\begin{figure}[t]
\setlength{\abovecaptionskip}{0.cm}
\setlength{\belowcaptionskip}{-0.3cm}
\captionsetup{font={footnotesize}}
\begin{center}
 \includegraphics[width=8.5cm]{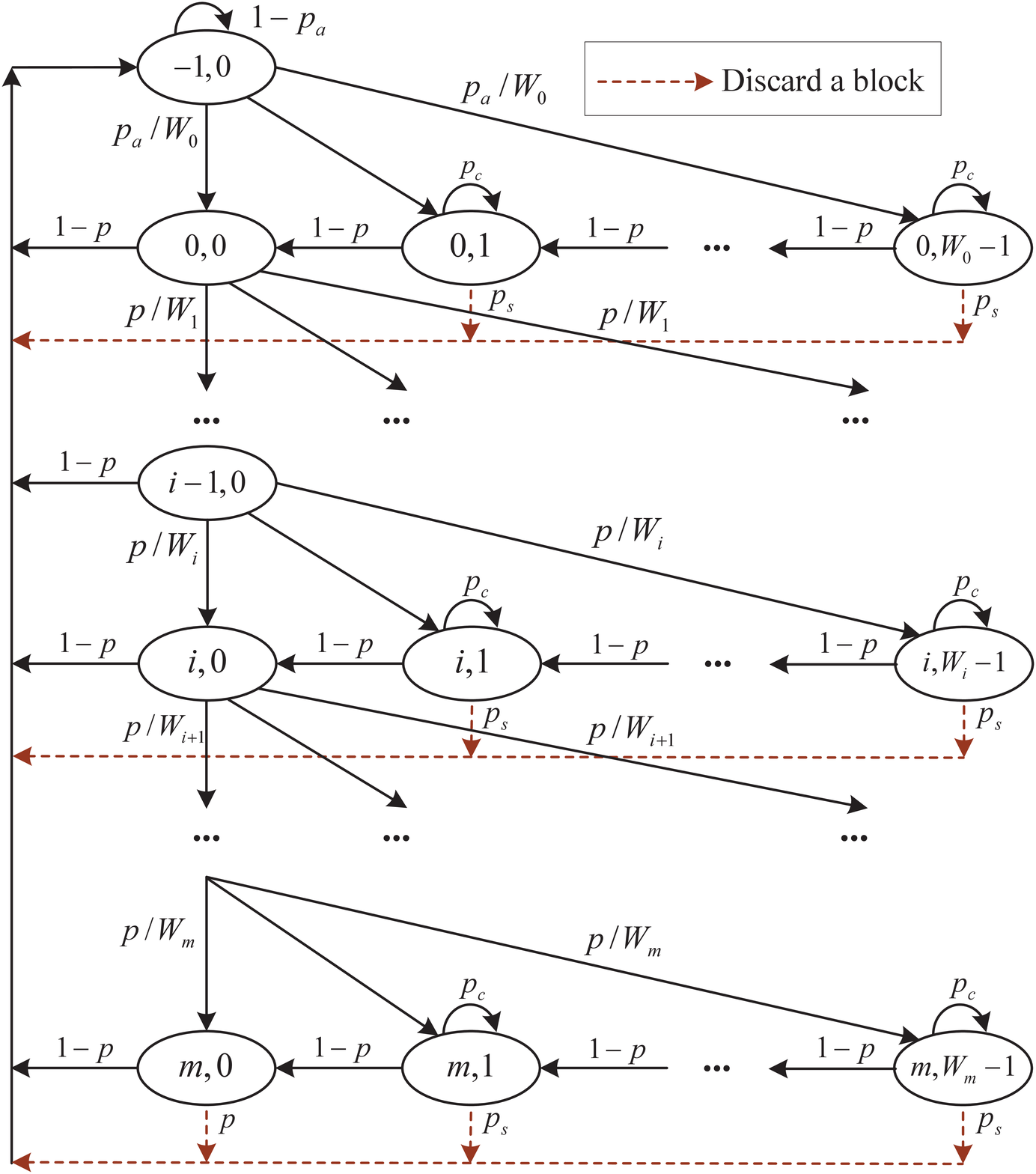}
 \end{center}
 \caption{{Markov chain model for BAC-3; When $p_{a}$ changes to $\tilde{p}_{a}$, this model is available for BAC-4.}}
\label{BAC-4}
\end{figure}

Since the BAC-3 and BAC-4 contain mining strategy II, a FN will pause mining when it enters the block backoff state until the new block is transmitted successfully or discarded. As a result, block queueing does not exist in the Markov chain model in Fig. \ref{BAC-4}, which is the main difference between the two Markov chain models.

\begin{bfseries}BAC-3\end{bfseries}: In no block and block backoff states, the expression of the one-step transition probabilities using BAC-3 are the same as BAC-1 in (\ref{onestep22}) and (\ref{onestep21}).
On the other hand, in block transmitting states, the one-step transition probabilities of BAC-3 are different with BAC-1 in (\ref{onestep23}). Since without block queueing in BAC-3, a FN must return to no block state after the FN transmits or discards its block. According to the analysis, the one-step transition probabilities of BAC-3 can be expressed as
\begin{equation}\label{onestepsecond}\small
\begin{split}
\begin{cases}
P\left\{ {-1,0\mid -1,0} \right\} \!=\! 1\!-\!p_{a},\\
P\left\{ {0,k\mid -1,0} \right\} \!=\! p_{a}/W_0,~~~~ \, ~~~~~~~~~~~~~~~~k\!\in\![0,W_0\!-\!1],\\
P\left\{ {-1,0 \mid i,k} \right\} \!=\! p_{s},~~~~~~ \, ~~~~i\!\in\![0,m],\,~~~~k\!\in\![1,W_i\!-\!1],\\
P\left\{ {i,k \mid i,k} \right\} \!=\! p_{c},~~~~~~~~~~~~~i\!\in\![0,m],\,~~~~k\!\in\![1,W_i\!-\!1],\\
P\left\{ {i,k\!-\!1 \mid i,k} \right\} \!=\! 1\!-\!p,~~~~~~i\!\in\![0,m],\,~~~~k\!\in\![1,W_i\!-\!1],\\
P\left\{ {i\!+\!1,k \mid i,0} \right\} \!=\! p/W_{i+1},~  ~i\!\in\![0,m\!-\!1],\,k\!\in\![0,W_i\!-\!1],\\
P\left\{ {-1,0 \mid i,0} \right\} \!=\! 1\!-\!p,~~~~~ \, ~~i\!\in\![0,m\!-\!1],\\
P\left\{ {-1,0 \mid m,0} \right\} \!=\! 1.
\end{cases}
\end{split}
\end{equation}

\begin{bfseries}BAC-4\end{bfseries}: The only difference between the BAC-4 and BAC-3 is the probability to leave no block state. Influenced by mining strategy I, the probability that a FN using BAC-4 leaves no block state during a step of Markov chain model is $\tilde{p}_{a}=(1\!-\!{p}_{s}\!-\!{p}_{c})[1\!-\!\exp (-\lambda\sigma)]$. So when $p_{a}$ changes to $\tilde{p}_{a}$, the one-step probabilities in (\ref{onestepsecond}) are available for BAC-4.

\subsection{Stationary Probabilities for BAC-3 and BAC-4}

Compared the chain regularities in Fig. \ref{Basic} with that in Fig. \ref{BAC-4}, we can know that (\ref{regular1}) can be directly applied to Fig. \ref{BAC-4}. Using (\ref{regular1}), $\pi _{i,0}$ can be expressed as
\begin{equation}\label{i0s}
\begin{split}
\pi _{i,0} \!=\! \frac{{1 \!-\! {{[(1 \!-\! p)/(1 \!-\! {p_c})]}^{{W_i}}}}}{{1 \!-\! (1 \!-\! p)/(1 \!-\! {p_c})}}\frac{p}{W_i}{\pi _{i - 1,0}} ,
\end{split}
\end{equation}
where $i\!\in\![1,m]$. For $i\!=\!0$, we use $p_{a}$ to replace $p$ in (\ref{i0s}) and obtain
\begin{equation}\label{00}
\begin{split}
\pi _{0,0} \!=\! \frac{{1 \!-\! {{[(1 \!-\! p)/(1 \!-\! {p_c})]}^{{W_0}}}}}{{1 \!-\! (1 \!-\! p)/(1 \!-\! {p_c})}}\frac{p_{a}}{W_0}{\pi _{-1,0}}.
\end{split}
\end{equation}
Now, using (\ref{i0s}) and (\ref{00}), a equation between no block state and transmitting states can be given by
\begin{equation}\label{mining1}
\begin{split}
\pi _{i,0} \!=\!f(i){\pi _{-1,0}},
\end{split}
\end{equation}
where {\small{$f(x)\!=\!\!{\prod\limits_{n = 0}^x \!{\frac{{1 - {{[(1 - p)/(1 - {p_c})]}^{{W_n}}}}}{{{W_n}}}} {{\left[\frac{p}{{1 - (1- p)/(1 - {p_c})}}\right]}^{x + 1}}} \frac{{{p_a}}}{p}$}} and $i\!\in\![0,m]$. Based on the chain regularity of no block state in Fig. \ref{BAC-4}, we obtain another equation between no block state and transmitting states as follows:
\begin{equation}\label{mining2}
\begin{split}
{\pi _{-1,0}} = &(1 \!-\! {p_a}){\pi _{-1,0}} \!+\! (1\! -\! p)\sum\limits_{i = 0}^m {{\pi _{i,0}}} \! +\! p \!\cdot\! {\pi _{m,0}} \! \\
&+\! {p_s}\left(1\! -\! \sum\limits_{i = 0}^m {{\pi _{i,0}}} \! -\! {\pi _{-1,0}}\right).
\end{split}
\end{equation}
By means of (\ref{mining1}), (\ref{mining2}) can be rewritten as
\begin{equation}\label{miningF}
{\pi _{-1,0}} \!=\! \frac{{{p_s}}}{{{p_a}\! + \!{p_s} \!-\! (1 \!- \!p \!-\! {p_s})\sum\limits_{i = 0}^m {f(i)} \!-\! pf(m)}}.
\end{equation}

Let the transmitting probability of BAC-3 be $\tau_{3}\!=\! \sum\limits_{i = 0}^m \!{{{\pi} _{i,0}}}$. Using (\ref{mining1}) and (\ref{miningF}), $\tau_{3}$ can be expressed as
\begin{equation}\label{ttao}
{\tau_{3}} \!=\! \frac{{{p_s}\sum\limits_{i = 0}^m {f(i)}}}{{{p_a}\! + \!{p_s} \!-\! (1 \!- \!p \!-\! {p_s})\sum\limits_{i = 0}^m {f(i)} \!-\! pf(m)}},
\end{equation}
where $p_{a}\!=\!(1\!-\!p_{s}\!-\!p_{c})[1-\exp(-\lambda\sigma)]$, ${p}=1-{(1-\tau_{3})}^{N-1}$, ${p}_{s}=(N-1){\tau_{3}}{(1-\tau_{3})}^{N-2}$, and ${p}_{c}={p}-{p}_{s}$. So $\tau_{3}$ is the only unknown parameter in (\ref{ttao}), which can be obtained through iteration method. Note that (\ref{ttao}) can be applied to BAC-4 when ${p}_{a}$ changes to $\tilde{p}_{a}$. To distinguish this difference between BAC-3 and BAC-4, we denote the transmitting probability of BAC-4 by $\tau_{4}$.

\section{Performance Analysis}

In the section, based on the stationary probabilities of Markov chain models, we analyse the closed-form expressions of the key performance metrics in B-WLAN by involving the impact of four BAC approaches.

\subsection{Transaction Throughput}

Transaction throughput is defined as the number of transactions that included by a valid block per second, i.e., transaction per second (tps). We denote transaction throughput by $\theta_{t}$. To calculating the $\theta_{t}$, we can multiply the number of blocks that successfully transmitted on the channel per second by the maximum number of transactions in a block. The maximum number of transactions in a block relates to the block size. To reduce the forking probability and achieve consensus in B-WLAN, we consider that a FN can transmit a full block after the backoff counter, which contains a block header and all related transactions. Let $s_b$ be the size of a block, $s_{h}$ be the size of the block header, $s_{t}$ be the average size of a transaction, $N_t$ be the maximum number of transactions in a block. Since each transmission contains a block header and all related transactions, we have $N_t\!=\!(s_{b}\!-\!s_{h})/s_{t}$. Based on the throughput analysis in \cite{3-performance}, the transaction throughput $\theta_{t}$ can be given by
\begin{equation}\label{Tthroughput}
\begin{split}
{\theta _{t}} \!=\! \frac{p_{1}N_t}{{{p_{0}}\sigma \! +\! {p_1}{T_s}\! +\! (1 \!-\! {p_0}\!-\!{p_1}){T_c}}},
\end{split}
\end{equation}
where $p_{0}=(1-\tau)^{N}$, $\tau\in\{\tau_{1},\tau_{2},\tau_{3},\tau_{4}\}$ representing the probability that the channel stays idle in a slot. $p_{1}=N\tau(1-\tau)^{N-1}$, $\tau\in\{\tau_{1},\tau_{2},\tau_{3},\tau_{4}\}$  representing the probability that a successful transmission occurs. Accordingly, $1 -{p_0}-{p_1}$ is the probability that a collision happens. $T_s$ is the average channel busy time when a successful transmission occurs. $T_c$ is the average channel busy time when a collision happens. According to \cite{3-performance}, $T_s$ and $T_c$ can be given by
\begin{equation}\label{TsTc}
\begin{split}
\begin{cases}
T_s \!=\! H \!+\! s_b\!+\!\rm{SIFS}\!+\! \delta \!+\!\rm{ACK}\!+\!\rm{DIFS}\!+\! \delta,\\
T_c \!=\! H \!+\! s_b\!+\!\rm{DIFS}\!+\! \delta,
\end{cases}
\end{split}
\end{equation}
where $H$ is the packet header; $\delta$ is the propagation delay; SIFS, DIFS and ACK denote the frame duration.

Note that by substituting $\tau\in\{\tau_{1},\tau_{2},\tau_{3},\tau_{4}\}$ into $p_{0}$ and $p_{1}$, the transaction throughput of BAC-1, BAC-2, BAC-3 and BAC-4 can be derived respectively using (\ref{Tthroughput}). For example, when $p_{0}=(1-\tau_{1})^{N}$ and $p_{1}=N\tau_{1}(1-\tau_{1})^{N-1}$, the result of (\ref{Tthroughput}) is the transaction throughput of BAC-1.

\subsection{Block Discard Rate}

Based on mining strategy, the way to derive block discard rate is different among four BAC approaches.

\begin{bfseries}BAC-1\end{bfseries}: Let $\theta_{d1}$ be the block discard rate of BAC-1 representing the number of forking blocks discarded by all FNs per second. Since BAC-1 do not contain mining strategy to pause mining, the FNs will keep mining all the time. In this case, $\theta_{d1}$ can be expressed as
\begin{equation}\label{Discard2}
\theta_{d1} \!=\! \lambda N\!-\!\theta_{s},
\end{equation}
where $\lambda N$ is the block generation rate of whole network. $\theta_{s}$ is the block successful transmission rate of whole network representing the number of blocks successfully transmitted by all FNs per second, which can be given by
\begin{equation}\label{Bthroughputb}
\begin{split}
{\theta _s} \!=\! \frac{{{p_{1}}}}{{{p_{0}}\sigma \! +\! {p_1}{T_s}\! +\! (1 \!-\! {p_0}\!-\!{p_1}){T_c}}}.
\end{split}
\end{equation}

\begin{bfseries}BAC-2\end{bfseries}: Let $\theta_{d2}$ be the block discard rate of BAC-2. According to mining strategy I, the FNs using BAC-2 should perform hash operations based on the channel condition. When the channel stays idle (probability $p_{0}$), all FNs will perform hash operations. When the channel contains a successful block transmission (probability $p_{1}$), only the FN in block transmitting states will perform hash operations. When the channel contains a collision (probability $1-p_{0}-p_{1}$), all the FNs in block transmitting states will perform hash operations. Based on the analysis, $\theta_{d2}$ can be given by
\begin{equation}\label{SIdiscardrate}\small
\begin{split}
{\theta_{d2}} \!=\! \frac{p_{0}N\lambda \sigma +p_{1}\lambda T_s+\sum\limits_{j = 2}^N \dbinom{N}{j} {{\tau ^j}{{(1 \!-\! \tau )}^{N - j}}}j\lambda T_c}{{{p_{0}}\sigma \! +\! {p_1}{T_s}\! +\! (1 \!-\! {p_0}\!-\!{p_1}){T_c}}}-\theta_{s},
\end{split}
\end{equation}
where $j$ is the expected number of FNs in block transmitting states during a collision, and the collision probability $1-p_{0}-p_{1}=\sum\limits_{j = 2}^N \dbinom{N}{j} {{\tau ^j}{{(1 \!-\! \tau )}^{N - j}}}$.

\begin{bfseries}BAC-3\end{bfseries}: Let $\theta_{d3}$ be the block discard rate of BAC-3. Instead of basing on channel condition, mining strategy II pauses mining based on the working states of a FN. So we use another way to study the block discard rate of BAC-3. The blocks will be discarded because of the following two cases. (i) When a successful transmission occurs on the channel, the FNs in no block state or block backoff states will discard blocks based on discard strategy. (ii) When a collision happens on the channel, the FNs in state $\{m,0\}$ will discard blocks based on CSMA/CA.

We first analyse the case (i). In B-WLAN, a successful transmission occurs with probability $p_{1}=N\tau(1-\tau)^{N-1}$, where $\tau$ is the transmitting probability and $N$ is the number of FNs.
Based on this expression, we can know that one of the FNs is in transmitting state and the other $N\!-\!1$ FNs are in no block state or block backoff states in this slot. The number of discarded blocks depends on how many FNs are in no block state and block backoff states. Using Binomial theorem \cite{7-course}, $(1\!-\!\tau)^{N-1}$ can be expanded as
\begin{equation}\label{expand1tao}
{(1 \!-\! \tau )^{N - 1}} \!=\! \sum\limits_{n_b = 0}^{N - 1}\dbinom{N \!-\!1}{n_b} {\pi _{-1,0}}^{N - 1 - n_b}\!\cdot\!{(1 \!-\! \tau  \!-\! {\pi _{-1,0}})^{n_b}},
\end{equation}
where $n_b$ is the number of FNs in block backoff states, $N-1-n_b$ is the number of FNs in no block state. Accordingly, we can know that the number of discarded blocks when a successful block transmission occurs is
\begin{equation}\label{ns}
n_{s}=n_b+(N-1-n_b)[1\!-\!\exp (-\lambda T_{s})].
\end{equation}
Using (\ref{expand1tao}) and (\ref{ns}) to calculate expected value, the block discard rate for case (i) can be given by
\begin{equation}\label{D1}
\begin{split}
{\theta _{ds}} \!=\! \frac{\sum\limits_{n_b = 0}^{N - 1} {n_s\!\cdot\!N\tau \!\dbinom{N \!-\!1}{n_b}\! {\pi _{-1,0}}^{N \!-\! 1 \!- n_b}\!\cdot{(1 \!-\! \tau  \!-\! {\pi _{-1,0}})^{n_b}} }}{{{p_{0}}\sigma \! +\! {p_1}{T_s}\! +\! (1 \!-\! {p_0}\!-\!{p_1}){T_c}}}.
\end{split}
\end{equation}

Now we analyse the case (ii). In B-WLAN, a collision happens with probability $1 \!-\! {p_0}\!-\!{p_1}\!=\!\sum\limits_{j = 2}^N \dbinom{N}{j} {{\tau ^j}{{(1 \!-\! \tau )}^{N - j}}}$. Based on this expression, we can know that $j$ FNs are in transmitting state and $N\!-\!j$ FNs are in no block state or block backoff states in this slot. The number of discarded blocks depends on how many FNs are in state $\{m,0\}$. So we use Binomial theorem to expand $\tau ^j$ as
\begin{equation}\label{expandtao}
{\tau ^j} \!=\! \sum\limits_{n_{c} = 0}^j \dbinom{j}{n_{c}} {{{(\tau \! -\! {\pi _{m,0}})}^{j - n_{c}}} \cdot } {\pi _{m,0}}^{n_{c}},
\end{equation}
where $j\!\in\![2,N]$. $n_c$ is the number of FNs in state $\{m,0\}$, and thus the number of discarded blocks when a collision happens is equal to $n_c$. Based on this analysis, the block discard rate for case (ii) can be given by
\begin{equation}\label{D2}
\begin{split}
{\theta _{dc}} \!=\! \frac{\sum\limits_{n_c = 0}^{j} {n_c\!\cdot\!\sum\limits_{j = 2}^N \!\dbinom{N}{j}\! {{(1 \!-\! \tau )}^{N \!- j}}\! \dbinom{j}{n_{c}}\!{(\tau \! -\! {\pi _{m,0}})^{j \!- n_{c}}} \cdot {\pi _{m,0}}^{n_{c}} }}{{{p_{0}}\sigma \! +\! {p_1}{T_s}\! +\! (1 \!-\! {p_0}\!-\!{p_1}){T_c}}}.
\end{split}
\end{equation}

Using (\ref{D1}) and (\ref{D2}), the block discard rate of BAC-3 is given by
\begin{equation}\label{D1D2}
\theta _{d3}\!=\!\theta _{ds}\!+\!\theta _{dc}.
\end{equation}

\begin{bfseries}BAC-4\end{bfseries}: Let $\theta_{d4}$ be the block discard rate of BAC-4. The difference between BAC-4 and BAC-3 is that the FN in no block state will pause mining during the block transmission, and thus the FN only performs hash operations during the channel idle time. So when a successful block transmission occurs in BAC-4, the FNs in no block state will not discard blocks. In this case, ${n_s}={n_b}$ for BAC-4, and we rewrite (\ref{D1}) as
\begin{equation}\label{D1F}
\begin{split}
{\tilde{\theta} _{ds}} \!=\! \frac{\sum\limits_{n_b = 0}^{N - 1} {n_b\!\cdot\!N\tau \!\dbinom{N \!-\!1}{n_b}\! {\pi _{-1,0}}^{N \!-\! 1 \!- n_b}\!\cdot{(1 \!-\! \tau  \!-\! {\pi _{-1,0}})^{n_b}} }}{{{p_{0}}\sigma \! +\! {p_1}{T_s}\! +\! (1 \!-\! {p_0}\!-\!{p_1}){T_c}}}.
\end{split}
\end{equation}
Using (\ref{D1F}) and (\ref{D2}), the block discard rate of BAC-4 is given by
\begin{equation}\label{D1D2F}
\theta _{d4}\!=\!\tilde{\theta} _{ds}\!+\!\theta _{dc}.
\end{equation}

\subsection{Block Utilization and Mining Pause Probability}

In blockchain network, when a block is transmitted successfully, it will be verified and stored by all FNs. The computational power included in this block will be used to enhance the security of the public ledger. To study how many blocks can be used for security, let $\eta$ be the block utilization in B-WLAN, defined as the (long-run) proportion of the blocks that is transmitted successfully.

Using block successful transmission rate ${\theta _s}$, the number of blocks successfully transmitted by FNs during the time period $T$ is ${\theta _s}T$. Using block discard rate ${\theta_d}\in\{\theta _{1},\theta _{2},\theta _{3},\theta _{4}\}$, we can obtain that the number of blocks discarded by FNs during the time period $T$ is ${\theta _d}T$. So the block utilization is given by
\begin{equation}\label{utilization}
\begin{split}
{\eta} \!=\!\lim \limits_{T \to \infty } \frac{{\theta _s}T }{{\theta _s}T\!+\!{\theta _d}T }\!=\!\frac{{\theta _s} }{{\theta _s}\!+\!{\theta _d} },
\end{split}
\end{equation}
where $\theta _s$ can be derived by (\ref{Bthroughputb}). ${\theta_d}\in\{\theta _{1},\theta _{2},\theta _{3},\theta _{4}\}$ can be derived by (\ref{Discard2}), (\ref{SIdiscardrate}), (\ref{D1D2}) and (\ref{D1D2F}), respectively.

Now, we study the stationary mining pause probability, which equals the (long-run) proportion of time that the mining is paused. Let this probability be $p_{m}$. Review that the average number of blocks generated by a FN per second is $\lambda\!=\!r/D$, where $r$ is the hashrate of a FN and $D$ is the hash difficulty. Without mining strategy, the whole network will generate average $\lambda NT$ blocks during the time period $T$. But in fact, because of the mining pause, the number of blocks generated by the whole network during $T$ decreases to $({\theta _s}\!+\!{\theta _d})T$. This yields
\begin{equation}\label{long-run}
\begin{split}
p_{m} \!=\!\lim \limits_{T \to \infty } \frac{\lambda NT\!-\!({\theta _s}\!+\!{\theta _d})T}{\lambda NT}\!=\!\frac{\lambda N\!-\!{\theta _s}\!-\!{\theta _d}}{\lambda N}.
\end{split}
\end{equation}

\section{Numerical Results and Discussions}

In this section, we use Matlab to calculate the closed-form expression of the performance metrics for comparing the performance of B-WLAN using BAC-1, BAC-2, BAC-3 and BAC-4, respectively.

\subsection{Parameter Settings}

We consider a B-WLAN with $N$ FNs. Based on \cite{3-performance}, the minimum contention window $W_{min}$ is set as $16$, the maximum contention window $W_{max}$ is set as $1024$, the maximum backoff stage $m$ is set as $6$, the packet header $H$ is set as $400$ bits, the size of ACK frame is set as $240$ bits, channel bit rate is set as $1$ Mbit/s, the propagation delay $\delta$ is set as $1$ $\mu s$, the time slot $\sigma$ is set as $50$ $\mu s$. Meanwhile, based on \cite{3-Mastering}, the size of the block header $s_h$ is set as $640$ bits, the size of a transaction $s_t$ is set as $2000$ bits.

%\begin{table}[htbp]
%\caption{Parameter Settings}
%\begin{tabular}{l|l}
%\hline
%\multicolumn{1}{l}{Parameter} & Value                                    \\
%\hline
%Minimum contention window $W_{min}$		    & $16$ \\
%Maximum contention window $W_{max}$         & $1024$ \\
%Maximum backoff stage $m$		            & $6$ \\
%MAC header                                  & $272$ bits \\
%PHY header                                  & $128$ bits \\
%The size of ACK frame    		            & $240$ bits \\
%Channel bit rate                            & $1$ Mbit/s \\
%Propagation delay $\delta$                  & $1$ $\mu s$ \\
%Time slot $\sigma$                          & $50$ $\mu s$ \\
%SIFS                                        & $28$ $\mu s$ \\
%DIFS                                        & $128$ $\mu s$ \\
%The size of block header $s_h$              & $642$ bits \\
%The size of a transaction $s_t$             & $2000$ bits \\
%\hline
%\end{tabular}
%\end{table}

\subsection{Performance Evaluation}

In the first experiment, we vary the number of transactions in a block $N_{t}$ from $1$ to $100$ to study the impact of $N_{t}$ on transaction throughput, block discard rate, block utilization and mining pause probability. Meanwhile, we set the number of FNs $N$ and the block generation rate $\lambda$ as $10$ and $50$ to conduct comparisons. The unit of $\lambda$ is block per second (bkps).

\begin{figure}[t]
\setlength{\abovecaptionskip}{0.cm}
\setlength{\belowcaptionskip}{-0.1cm}
\captionsetup{font={footnotesize}}
\begin{center}
\includegraphics[width=7cm]{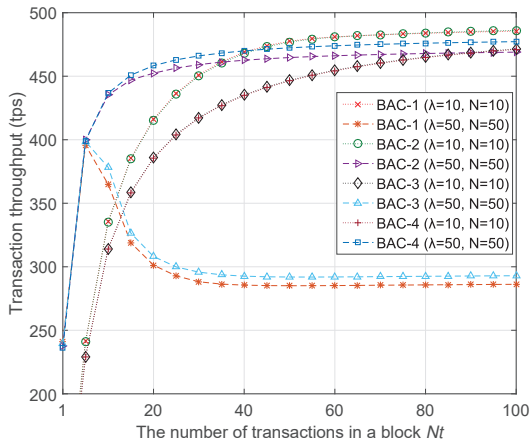}
\end{center}
\caption{Transaction throughput vs. $N_{t}$}%; the lowest value shown is $10^{-6}$}
\label{TPSvsN}
\end{figure}

The transaction throughput of BAC-1, BAC-2, BAC-3 and BAC-4 are derived by substituting the stationary probabilities $\tau_{1}$, $\tau_{2}$, $\tau_{3}$ and $\tau_{4}$ into (\ref{Tthroughput}) respectively. \begin{bfseries}BAC-1\end{bfseries}: Fig. \ref{TPSvsN} shows that when $\lambda\!=\!10$, $N\!=\!10$, the BAC-1 can help B-WLAN to achieve a high transaction throughput, i.e., up to 480 tps in ideal channel condition. With the increase of $N_{t}$, one block can contain more transactions after the backoff counter and the backoff delay of a single transaction decreases, so the transaction throughput quickly increases at the beginning. But at the same time, due to limited channel resource, the increasing rate declines gradually, and the throughput reaches a upper bound. When $\lambda\!=\!50$ and $N\!=\!50$, the transaction throughput of BAC-1 begin to decrease after $N_{t}=5$ and cannot reach the upper bound. Because when $\lambda$ and $N$ is large, there are too many blocks occurring in the backoff states and the collisions cannot be effectively avoided by the backoff counter. \begin{bfseries}BAC-2\end{bfseries}: Compared with BAC-1, BAC-2 can help B-WLAN to reach the throughput upper bound, no matter $\lambda\!=\!10$, $N\!=\!10$ or $\lambda\!=\!50$, $N\!=\!50$. This phenomenon reflects that mining strategy I can effectively pause mining to balance the block generation rate. This action helps the backoff scheme to avoid collisions and maintain a high transaction throughput in high load case. \begin{bfseries}BAC-3\end{bfseries}: The curves under BAC-3 is similar to that under BAC-1, while BAC-1 behaves sightly better in small $\lambda$ and $N$, and BAC-3 behaves better in large $\lambda$ and $N$. This means strategy II is suitable for high block generation rate case. \begin{bfseries}BAC-4\end{bfseries}: It is shown that BAC-4 has the highest throughput when $\lambda\!=\!50$, $N\!=\!50$. So BAC-4 is the best choice in this high load case. For $\lambda\!=\!10$, $N\!=\!10$, BAC-1 and BAC-2 have a very similar curve, but BAC-2 is a better choice in low load case since the mining strategy in it saves power.

\begin{figure}[t]
\setlength{\abovecaptionskip}{0.cm}
\setlength{\belowcaptionskip}{-0.3cm}
\captionsetup{font={footnotesize}}
\begin{center}
\includegraphics[width=7cm]{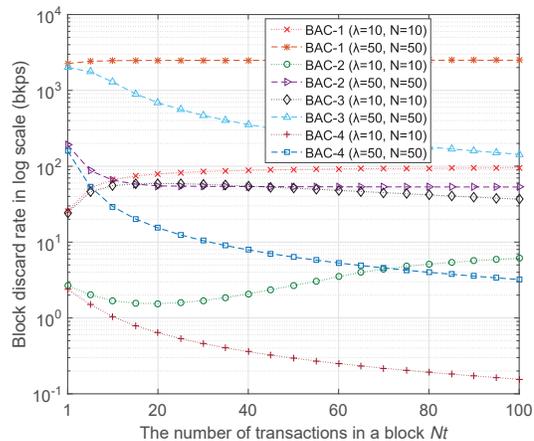}
\end{center}
\caption{Block discard rate (log scale) vs. $N_{t}$}%; the lowest value shown is $10^{-6}$}
\label{DRvsN}
\end{figure}

The block discard rate of BAC-1, BAC-2, BAC-3 and BAC-4 are derived by substituting $\tau_{1}$, $\tau_{2}$, $\tau_{3}$ and $\tau_{4}$ into (\ref{Discard2}), (\ref{SIdiscardrate}), (\ref{D1D2}) and (\ref{D1D2F}) respectively. \begin{bfseries}BAC-1\end{bfseries}: Fig. \ref{DRvsN} shows that when $\lambda\!=\!10$, $N\!=\!10$ and $\lambda\!=\!50$, $N\!=\!50$, the block discard rate of BAC-1 always increases with $N_{t}$. Because a larger $N_{t}$ means a longer block transmission time, and BAC-1 does not pause mining during the transmission time. So more forking blocks are generated and discarded. \begin{bfseries}BAC-2\end{bfseries}: It can be noticed that when $\lambda\!=\!10$, $N\!=\!10$ the block discard rate firstly decreases and then increases with $N_{t}$. The reason is that with the increase of $N_{t}$, the block transmission time becomes longer, which increases the mining pause time and reduces the forking probability. So the discard rate decreases with $N_{t}$ at first. But on the other hand, a longer block transmission time also incurs more queuing blocks. A large number of queuing blocks cause a high forking probability and cannot be well addressed by mining pause. So the discard rate increases finally. When $\lambda\!=\!50$, $N\!=\!50$, the block discard rate of BAC-2 always decrease with $N_{t}$. This is because a large $\lambda\!=\!50$ and $N\!=\!50$ results in much collision on the channel, and mining strategy I frequently pause mining to reduces the forking blocks. \begin{bfseries}BAC-3\end{bfseries}: It is shown that when $\lambda\!=\!10$, $N\!=\!10$ the block discard rate firstly increases and then decreases with $N_{t}$. Because the FN using BAC-3 keeps mining during the block transmission of other FNs, more forking blocks are discarded after a block transmission. But on the other hand, with a much longer transmission time, the impact of mining pause significantly increases, which slows down the total block generation rate and reduces forking probability. For $\lambda\!=\!50$, $N\!=\!50$, the block discard rate always decrease with $N_{t}$ due to high block generation rate prolonging the average mining pause time. \begin{bfseries}BAC-4\end{bfseries}: The block discard rate always decreases with $N_{t}$. For a given $\lambda$ and $N$, the block discard rate of BAC-4 is lower than the other BAC approaches. This means mining strategy I and strategy II are both effective to reduce the block discard rate.

\begin{figure}[t]
\captionsetup{font={footnotesize}}
\setlength{\abovecaptionskip}{0.cm}
\setlength{\belowcaptionskip}{-0.3cm}
\begin{center}
\includegraphics[width=7cm]{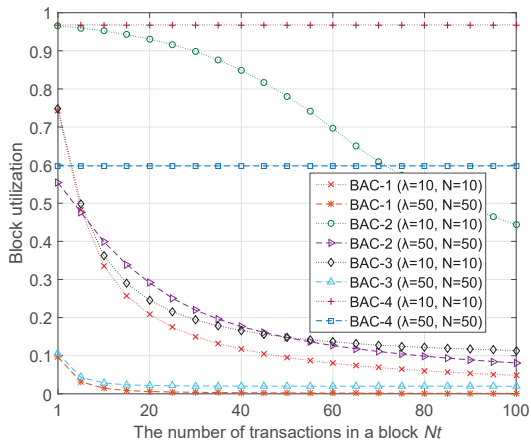}
\end{center}
\caption{Block utilization vs. $N_{t}$}%; the lowest value shown is $10^{-6}$}
\label{UvsN}
\end{figure}

The block utilization of BAC-1, BAC-2, BAC-3 and BAC-4 are derived by substituting $\tau_{1}$, $\tau_{2}$, $\tau_{3}$ and $\tau_{4}$ into (\ref{utilization}) respectively. \begin{bfseries}BAC-1\end{bfseries}: Fig. \ref{UvsN} shows that the block utilization of BAC-1 always decease with $N_{t}$. Because a large $N_{t}$ prolongs block transmission time, and much computational power of FNs is wasted before receiving the new block, which results in a low block utilization. \begin{bfseries}BAC-2\end{bfseries}: The block utilization also decease with $N_{t}$. But for a given $N_{t}$, the block utilization of BAC-2 is higher than BAC-1, since the mining strategy I in it reduces forking probability. \begin{bfseries}BAC-3\end{bfseries}: For a given $N_{t}$, it is shown that the block utilization of BAC-3 is lower than that of BAC-2. This phenomenon reflects that mining strategy I has a better effect on block utilization, compared with strategy II. The reason is that strategy I pauses mining based on the detection of block transmission which happens more frequently than strategy II. But on the other hand, the implement cost of strategy I will be higher than strategy II. \begin{bfseries}BAC-4\end{bfseries}: It is observed that the block utilization of BAC-4 is not affected by $N_{t}$ and stabilize at a high level. This important results indicates that two mining strategies can work well together to achieve a optimal block utilization.

\begin{figure}[t]
\captionsetup{font={footnotesize}}
\begin{center}
\includegraphics[width=7cm]{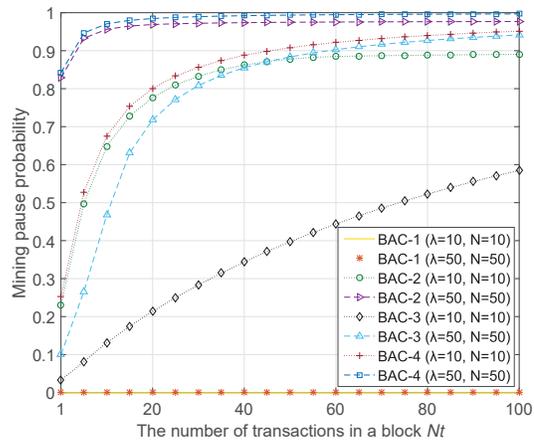}
\end{center}
\caption{Mining pause probability vs. $N_{t}$}%; the lowest value shown is $10^{-6}$}
\label{SPvsN}
\end{figure}

The mining pause probability of BAC-1, BAC-2, BAC-3 and BAC-4 are derived by substituting $\tau_{1}$, $\tau_{2}$, $\tau_{3}$ and $\tau_{4}$ into (\ref{long-run}) respectively. \begin{bfseries}BAC-1\end{bfseries}: BAC-1 does not contain mining strategy, so the mining pause probability is $0$. \begin{bfseries}BAC-2\end{bfseries}: Fig. \ref{SPvsN} shows that the mining pause probability of BAC-2 increases with $N_{t}$. Because a larger $N_{t}$ needs a longer transmission time, it increases mining pause time. \begin{bfseries}BAC-3\end{bfseries}: The mining pause probability of BAC-3 is lower than that of BAC-2. This is because strategy I pauses mining whenever a block is transmitted, and mining strategy II pauses mining when a FN transmit its own block. For example, when the channel contains a block transmission, there are $N-1$ FNs pausing mining in BAC-2, but there is only one FN pausing mining in BAC-3. \begin{bfseries}BAC-4\end{bfseries}: For a given $N_{t}$, BAC-4 has a highest mining pause probability, which means it saves more computational power than other BAC approaches.

In the second experiment, we vary the block generation rate $\lambda$ at each FN from $1$ to $100$ to compare the performance of four BAC approaches with the number of FNs. Meanwhile, we set $N_{t}=10$ in this experiment, and one can capture the impact of $N_{t}$ based on the result of first experiment.

\begin{figure}[t]
\captionsetup{font={footnotesize}}
\setlength{\abovecaptionskip}{0.cm}
\setlength{\belowcaptionskip}{-0.3cm}
\begin{center}
\includegraphics[width=7cm]{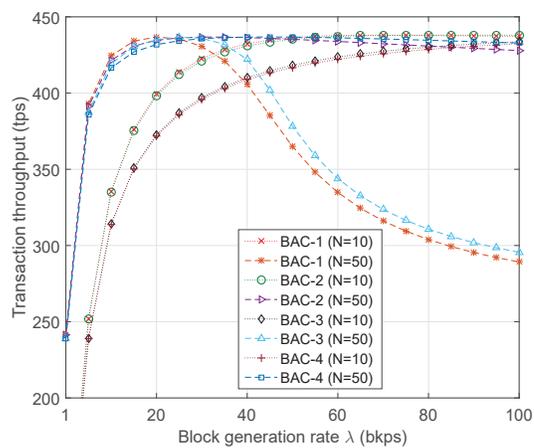}
\end{center}
\caption{Transaction throughput vs. $\lambda$}%; the lowest value shown is $10^{-6}$}
\label{TPSvsL}
\end{figure}

\begin{bfseries}BAC-1\end{bfseries}: Fig. \ref{TPSvsL} shows that the transaction throughput of BAC-1 increases with block generation rate $\lambda$ before reaching the maximum value. Because when $\lambda$ is low there are very few blocks transmitting on the channel, the collision probability is low and the transaction throughput increases quickly with $\lambda$. With the increases of $\lambda$, the collisions are addressed by the CSMA/CA scheme and thus the transaction throughput reaches the maximum value. After that, the transaction throughput decreases with $\lambda$ due to too many blocks occurring in the backoff states and the collisions cannot be effectively avoided by the backoff counter. \begin{bfseries}BAC-2\end{bfseries}: When $N=10$, the BAC-2 has a similar throughput with BAC-1. However, BAC-1 achieves this throughput by using more computational power, while BAC-2 achieves this throughput with a higher implement cost (block detection). When $N=50$, the transaction throughput of BAC-2 decreases much slower than BAC-1 after reaching the maximum value. Because BAC-2 contains mining strategy I to pause mining and slow down $\lambda$ in high load case, which reduces collision probability and maintains a high throughput. \begin{bfseries}BAC-3\end{bfseries}: BAC-3 is suitable for the case when four BAC approaches have a similar throughput, e.g., $\lambda=100$, $N=10$ or $\lambda=20$, $N=50$. This is because the implement of mining strategy II is much easier than strategy I (based on the principles in section III). So for implement cost, BAC-1 is similar to BAC-3, which are lower than BAC-2 and BAC-4. Meanwhile, the power consumption of BAC-1 is higher than BAC-3. \begin{bfseries}BAC-4\end{bfseries}: It is shown that the curve of BAC-4 is similar to BAC-2 when $N=50$, while BAC-4 has a lower power consumption. So BAC-4 is a better choice for an optimal throughput with large $\lambda$ and $N$, e.g., $\lambda=100$ and $N=50$.

\begin{figure}[t]
\captionsetup{font={footnotesize}}
\setlength{\abovecaptionskip}{0.cm}
\setlength{\belowcaptionskip}{-0.3cm}
\begin{center}
\includegraphics[width=7cm]{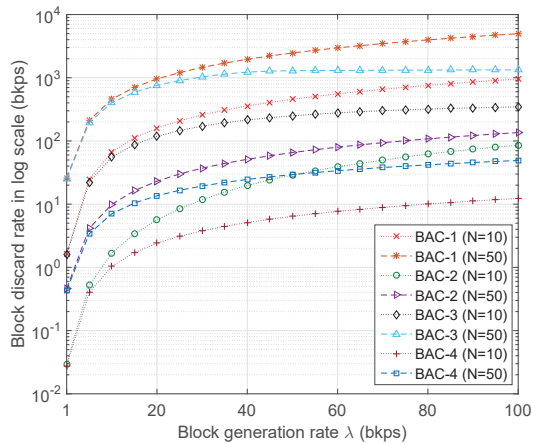}
\end{center}
\caption{Block discard rate (log scale) vs. $\lambda$}%; the lowest value shown is $10^{-6}$}
\label{DRvsL}
\end{figure}

Fig. \ref{DRvsL} shows that the block discard rate of four BAC approaches increases with $\lambda$. Because a larger $\lambda$ results in more blocks simultaneously generated in the backoff state. In this case, only one block can be transmitted successfully, and the other blocks are discarded due to forking. So the block discard rate increases with $\lambda$. Meanwhile, it can be observed that the block discard rate of BAC-1 is higher than the other BAC approaches. This means mining strategy can reduce forking blocks effectively, especially in the scenario with high block generation rate.

\begin{figure}[htbp]
\captionsetup{font={footnotesize}}
\begin{center}
\includegraphics[width=7cm]{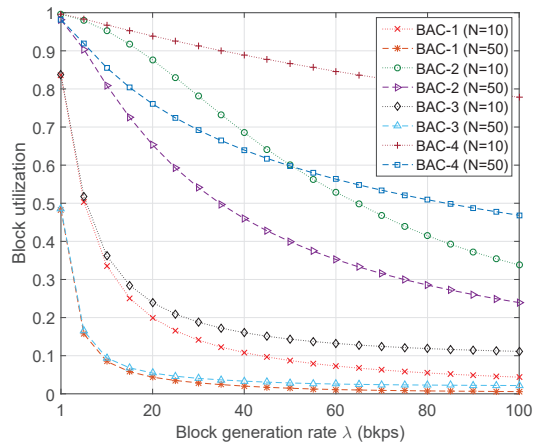}
\end{center}
\caption{Block utilization vs. $\lambda$}%; the lowest value shown is $10^{-6}$}
\label{UvsL}
\end{figure}

Fig. \ref{UvsL} shows that the block utilization of four BAC approaches always decrease with $\lambda$. When block utilization is lower than $0.5$, most blocks generated by FNs are discarded due to forking. The reason is that when multiple blocks simultaneously generated in the backoff states, the discard probability for a single block is very high. Since the forking blocks cannot protect the main chain of the ledger, a low block utilization ``dilutes" the computational power of FN and affects the security of blockchain network. In addition, Fig. \ref{TPSvsL} and Fig. \ref{UvsL} demonstrate the trade-off between transaction throughput and block utilization. This means one can adjust PoW hash difficulty to accelerate block generation rate for achieving a high transaction throughput, but at the same time the block utilization decreases.

\begin{figure}[t]
\captionsetup{font={footnotesize}}
\setlength{\abovecaptionskip}{0.cm}
\setlength{\belowcaptionskip}{-0.3cm}
\begin{center}
\includegraphics[width=7cm]{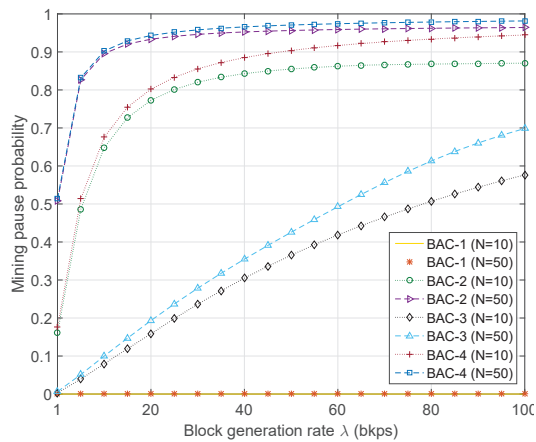}
\end{center}
\caption{Mining pause probability vs. $\lambda$}
\label{SPvsL}
\end{figure}

Fig. \ref{SPvsL} shows that the mining pause probability of BAC-2, BAC-3 and BAC-4 always increase with $\lambda$. Meanwhile, for a given $N$, it is shown that the mining pause probability of BAC-4 is the highest, and the mining pause probability of BAC-2 is higher than that of BAC-3. Note that the mining pause probability affects the computational power consumption of FNs. So the results in Fig. \ref{UvsL} and Fig. \ref{SPvsL} prove that mining strategy can save the computational power while improving block utilization of B-WLAN.

Note that the performance comparisons in this work refer to four BAC approaches that contain different strategies. For the baseline approach without any strategies, it can be predicted that the performance would be worse than BAC-1, since the forking blocks in queue consume much channel resource and increase block propagation delay.

\section{Related Work}

In recent years, many researches have been carried out to improve the transaction throughput of PoW-based blockchain. Bitcoin-NG \cite{19-Bitcoin-NG} selects a leader to post multiple blocks, thus increasing the block generation rate of PoW for improving the transaction throughput. Hybrid-IoT \cite{19-Hybrid-iot} proposes a two-tier blockchain architecture, where subgroups of full nodes achieve consensus through PoW algorithm and the connection among the sub-blockchains employs a Byzantine fault-tolerant framework. Monoxide \cite{19-Monoxide} adopts multiple independent and parallel PoW sub-blockchains termed as zones, in which different zones can conduct trading using the cross-zone algorithm. Tangle \cite{9-tangle} uses a directed acyclic graph (DAG)-based ledger to replace the conventional single chain-based ledger. After solving a simple PoW task, the nodes can insert their blocks into DAG-based ledger at any time, which result in a much higher throughput. Although the high throughput can be achieved by these new blockchain systems, the security is compromised since generating multiple sub-blockchains ``dilutes" the mining power of honest nodes. Without generating sub-blockchains, this paper proposes a discard strategy to address the blockchain forking problem, which act as a key enabler to accelerate block generation rate and improve transaction throughput. The discard strategy is proposed by considering the impact of CSMA/CA channel contention on blockchain consensus process, and has not been studied in previous work.

The high computational power consumption is also a serious problem in PoW-based blockchain. To address this problem, the authors in \cite{10-auction} propose an auction-based market model to offload the PoW computational tasks to the cloud/fog computing server. Two bidding schemes, i.e, constant-demand scheme and multi-demand scheme are considered to maximize the social welfare of the blockchain network. In \cite{11-Multi-Leader}, the authors study the offloading from miners to cloud/edge servers using a multi-leader multi-follower game. Based on alternating direction method of multipliers (ADMM) algorithm, the utilities of miners and the profits of servers are jointly optimized. Considering the game theory cannot deal with the dynamics of the wireless environment, the authors in \cite{12-NOMA} propose a multi-agent deep reinforcement learning (DRL) approach to minimize the long-term cost of the PoW task offloading. The above work addresses power consumption of PoW based on offloading model. However, PoW task offloading cannot save the overall computational power, since the server becomes a substitute for miner. Different from offloading approaches, we study the relationship between block transmission delay and forking problem, then propose mining strategy to pause mining during block transmission, which reduces the meaningless computational power consumption on forking blocks. %Using mining strategy to reduce the meaningless computational power consumption on forking blocks have not been considered in previous work.

To validate the effectiveness of the proposed approach, mathematical models are required to quantitatively study the performance of blockchain. In \cite{20-DAGsecurity}, the authors present Markov chain model to analyse the consensus process of DAG-based ledger. Using transition probabilities, the authors derive blockchain performance metrics, i.e., cumulative weight increasing rate and consensus delay. In \cite{20-raft}, the authors uses Markov chain model to analyse the performance of Raft consensus algorithm. The impact of packet loss rate, election timeout, and network size on Raft network split probability have been derived. Except Markov chain models, the authors in \cite{20-Blockchain-enabled} employ a signal-to-interference-plus-noise ratio (SINR) model to analyse the transaction transmission successful rate and transaction throughput of PoW-based blockchain. An optimal node deployment algorithm has been designed to maximize transaction throughput. However, the above models do not study the impact of communication protocol (e.g. CSMA/CA) on blockchain performance.
%In \cite{16-Stochastic}, the authors develop a stochastic model to analyze the impact of the block transmission delay and hashrate of full nodes on Blockchain performance in terms of the block generation rate and the required computational power for launching a successful attack.
In \cite{20-DAG-CSMA}, the authors develop a queuing model to analyze the impact of CSMA/CA-based transmission delay on the consensus efficiency of DAG-based ledger. But the CSMA/CA-based transmission delay is considered as an average value to increase confirmation delay, and the authors do not study how the random backoff scheme in CSMA/CA affects the block transmission. In view of this, this paper extends the Markov chain model in \cite{3-performance} to study how the proposed strategies and random backoff scheme affect the performance of blockchain. To the best of our knowledge, the transaction throughput has not been studied by involving blockchain strategy and the channel contention of CSMA/CA simultaneously. Meanwhile, the block discard rate, block utilization and mining pause probability are the key performance metrics to show the effectiveness of the proposed strategies, which has not been analysed in previous work.

\section{Conclusions and Future work}

In this work, we propose mining strategies and a discard strategy to reduce the meaningless computational power consumption on forking blocks and improve the transaction throughput in B-WLAN. Based on the proposed strategies, we design four BAC approaches and use Markov chain models to conduct the performance comparisons. Calculation results show that the discard strategy in BAC can help B-WLAN to achieve a high transaction throughput, which could meet the needs of the massive service requests in next-generation wireless network. Meanwhile, the mining strategy can pause the mining of FN to reduce forking probability effectively, which both saving the computational power and improving block utilization. In addition, it is shown that the block size (related to the number of transactions in a block) and PoW hash difficulty (related to block generation rate) will affect the trade-off between transaction throughput and block utilization, which can provide an analytical guideline for building a optimal and secure B-WLAN in the future.

As future work, we will study the performance of baseline approach without any strategies. The analysis of forking probability in queuing and backoff process is the main challenge to derive the transaction throughput and block utilization.

\appendix[Proof of $T_{q}$ in Equation (\ref{Tq})]

$T_{q}$ is defined as the expected time of a block spent on backoff and transmitting states counting from the block generation to the successful transmission, which can be given by
\begin{equation}\label{Tqprove}\small
T_{q}\!=\! \sum\limits_{i = 0}^m {p_e}(i)\left[(i{T_c}\! +\! {T_s})\!+\!{\sum\limits_{n = 0}^i {\frac{{{W_n} \!-\! 1}}{2}} }\left(\sigma\!+\!\frac{{p}_{c}}{1\!-\!{p}}T_{c}\right)\right].
\end{equation}

\begin{bfseries}Proof\end{bfseries}: According to the definition of $T_{q}$, we have
\begin{equation}\label{Tqp}\small
T_{q}\!=\!T_{b}\!+\!T_{t},
\end{equation}
where $T_{b}$ and $T_{t}$ are defined as the expected time spent on backoff states and transmitting states respectively.

Now we analyse $T_{b}$ and $T_{t}$. After a block enters the exponential backoff scheme, it can exit the backoff scheme either through a successful transmission in block transmitting states $\{i,0\}$ ($i\!\in\![0,m]$) or through a block discard (the dotted lines in Fig. \ref{Basic}). If the block exits the backoff scheme through a block discard, the queue of the FN will be cleared and thus the time spent on backoff and transmitting states should be considered as $0$ when we calculate the expected value. Therefore, to derive $T_{b}$ and $T_{t}$, we can only calculate the case when the block exits the backoff scheme through a successful transmission in transmitting states. Let $p_{e}(i)$ ($i\!\in\![0,m]$) be the probability that a block exits the backoff scheme through a successful transmission in state $\{i,0\}$. In any backoff state, the probability that a block is not discarded can be given by
\begin{equation}\label{anystate}\small
(1\!-\!p)\!+\!(1\!-\!p)p_{c}\!+\!(1\!-\!p)p_{c}^{2}\! +\! \cdots \!+(1\!-\!{p}){p}_{c}^{\infty}\!=\!\frac{1\!-\!{p}}{1\!-\!{p}_{c}}.
\end{equation}
Using (\ref{anystate}), the probability that a block is not discarded during backoff stage $i$ ($i\!\in\![0,m]$) can be expressed as
\begin{equation}\label{statei}\small
\begin{split}
&\frac{1}{{{W_i}}}{(\frac{{1 \!-\! {p}}}{{1 \!-\! {{p}_c}}})^0} \!+\! \frac{1}{{{W_i}}}{(\frac{{1 \!-\! {p}}}{{1 \!-\! {{p}_c}}})^1} \!+\!  \cdots \! +\! \frac{1}{{{W_i}}}{(\frac{{1 \!-\! {p}}}{{1 \!- \!{{p}_c}}})^{{W_i} - 1}} \!\\
= & \frac{1}{{{W_i}}}\frac{{1 \!-\! {{[(1 \!- \!{p})/(1 \!- \!{{p}_c})]}^{{W_i}}}}}{{1 \!-\! (1\! -\! {p})/(1 \!-\! {{p}_c})}}.
\end{split}
\end{equation}
By means of (\ref{statei}), the probability that a block exits the backoff scheme through a successful transmission in state $\{i,0\}$ is
\begin{equation}\label{pe}\small
{p_e}(i) \!=\! \prod\limits_{n = 0}^i \!{\frac{{1 \!-\! {{[(1 \!-\! {p})/(1\! -\! {{p}_c})]}^{{W_n}}}}}{{{W_n}}}}\left[\frac{{p}}{{1 \!-\! (1\! - \!{p})/(1 \!-\! {{p}_c})}}\right ]^{i+\!1}\frac{1 \!-\! {p}}{{p}}.
\end{equation}
So we obtain the expected time spent on transmitting states is
\begin{equation}\label{Tt}
\begin{split}
{T_t} \!&= \!{p_e}(0){T_s} \!+\! {p_e}(1)({T_c} \!+\! {T_s}) \!+\!  \cdots  \!+ \!{p_e}(m)(m{T_c}\! + \!{T_s})\\
\!&=\!\sum\limits_{i = 0}^m {{p_e}(i)} (i{T_c}\! +\! {T_s}).
\end{split}
\end{equation}

For $T_{b}$, there are too many possible paths to go through backoff states in Fig. \ref{Basic}, and thus it is very difficult to directly calculate the expected time spent on backoff states. Therefore, we use average value to estimate $T_{b}$. Let $\overline{W}$ be the average number of backoff windows, and $T_{w}$ be the average time spent on each backoff window. By estimation, $T_{b}$ is given by
\begin{equation}\label{Tb}
T_{b}\!\approx\!\overline{W}\!\cdot\!T_{w},
\end{equation}
where $\overline{W}$ can be given by
\begin{equation}\label{W}\small
\begin{split}
\overline W  \!&=\! {p_e}(0)\frac{{{W_0} \!- \!1}}{2}\! + \!{p_e}(1)\sum\limits_{n = 0}^1 {\frac{{{W_n}\! - \!1}}{2}} \! + \! \cdots  \!+ \!{p_e}(m)\sum\limits_{n = 0}^m {\frac{{{W_n} \!-\! 1}}{2}}\\
\!& =\! \sum\limits_{i = 0}^m {\sum\limits_{n = 0}^i {{p_e}(i)\frac{{{W_n} \!-\! 1}}{2}} },
\end{split}
\end{equation}
and $T_{w}$ can be expressed as
\begin{equation}\label{Tw}\small
T_{w}\!=\!\sigma\!+\!\frac{{p}_{c}}{1\!-\!{p}}T_{c}.
\end{equation}
By means of (\ref{W}) and (\ref{Tw}), (\ref{Tb}) rewrites as
\begin{equation}\label{Tb1}\small
T_{b}\!\approx\!\sum\limits_{i = 0}^m {\sum\limits_{n = 0}^i {{p_e}(i)\frac{{{W_n} \!-\! 1}}{2}} }\left(\sigma\!+\!\frac{{p}_{c}}{1\!-\!{p}}T_{c}\right).
\end{equation}
After we have $T_{t}$ in (\ref{Tt}) and $T_{b}$ in (\ref{Tb1}), $T_{q}$ can be expressed as (\ref{Tqprove}).

\end{document}